\scrollmode

\documentclass[12pt,twoside,a4paper]{article}
\usepackage[dvips]{epsfig}
\newcommand{\al}{\alpha}

\newcommand{\g}{\gamma}
\newcommand{\G}{\Gamma}
\newcommand{\de}{\delta}

\newcommand{\e}{\epsilon}

\newcommand{\si}{\sigma}

\newcommand{\simgt}{\,\rlap{\lower 3.5 pt \hbox{$\mathchar \sim$}} \raise 1pt
 \hbox {$>$}\,}
\newcommand{\simlt}{\,\rlap{\lower 3.5 pt \hbox{$\mathchar \sim$}} \raise 1pt
 \hbox {$<$}\,}

\newcommand{\equ}[2]{\begin{equation} \label{#1} #2 \end{equation} }

\begin{document}
\thispagestyle{empty}
\title{\vskip-3cm{\baselineskip14pt
\centerline{\normalsize DESY 97--039\hfill ISSN 0418--9833}
\centerline{\normalsize hep--ph/9703302\hfill}
\centerline{\normalsize March 1997\hfill}}
\vskip1.5cm
Inclusive Jet Production with Virtual Photons \\ in Next-to-Leading 
Order QCD \\
\author{M.~Klasen$^1$,  G.~Kramer$^2$, B.~P\"otter$^2$ \\
$^1$ Deutsches Elektronen-Synchrotron DESY, \\
Notkestr. 85, D-22607 Hamburg, Germany, \\
e-mail: klasen@mail.desy.de \\
$^2$ II. Institut f\"ur Theoretische Physik\thanks{Supported
by Bundesministerium f\"ur Forschung und Technologie, Bonn, Germany,
under Contract 05~7~HH~92P~(0),
and by EEC Program {\it Human Capital and Mobility} through Network
{\it Physics at High Energy Colliders} under Contract
CHRX--CT93--0357 (DG12 COMA).},
Universit\"at Hamburg\\
Luruper Chaussee 149, D-22761 Hamburg, Germany\\
e-mail: kramer@mail.desy.de, poetter@mail.desy.de} }
\date{}
\maketitle
\begin{abstract}
\medskip
\noindent
We present a next-to-leading order calculation for the virtual
photoproduction of one and two jets in $ep$ collisions. Soft and collinear
singularities are extracted using the phase space slicing method. The
collinear photon initial state singularity depends logarithmically on
the mass of the virtual photon and is absorbed into the virtual
photon structure function. An $\overline{\rm MS}$ factorization
scheme is defined similarly to the real photon case. Numerical results
are presented for HERA conditions using the Snowmass jet definition
for inclusive single jet and dijet cross sections. We study the
dependence of these cross sections on the transverse energies and
rapidities of the jets. Finally, we compare the ratio of the
experimentally defined resolved and direct cross sections 
with recent ZEUS data as a function of the photon virtuality $P^2$.
\end{abstract}

\newpage
\thispagestyle{empty}
\vspace*{12cm}

\newpage
%************************************************************
%************************************************************
\section{Introduction}

In $ep$ scattering at HERA interactions between photons of small
virtuality $P^2$ and protons produce jets of high transverse energy $E_T$.
The presence of this hard scale $E_T$ allows the application of
perturbative QCD to predict cross sections for the production of
two or more high-$E_T$ jets which can be confronted with experimental
data. This offers the opportunity to test QCD and to constrain
the structure of the colliding particles.

At leading order (LO) QCD two distinct processes are responsible
for the production of jets. In the direct photon process, the
photon interacts as a point-like object with a parton in the proton,
whereas in the resolved process the photon acts as a source of
partons which then scatter with the partons coming from the proton.
Both LO processes are characterized by having two outgoing jets
of large transverse energy. Studies of dijet photoproduction at
HERA have shown that both classes of processes are present for the
case of quasi-real photons, i.e. photons of extremely small
virtuality  $P^2\simeq 0$ \cite{1, 2}. The comparison of theoretical
predictions \cite{3, 4} with these \cite{2} and more recent
experimental data \cite{5} have given us some confidence that the
parton distribution functions for the real photon available in
the literature \cite{6} are consistent with the dijet production data.
The parton content of photons with virtuality $P^2 = 0$ is reasonably
constrained by data from deep inelastic  scattering \cite{7}.
Unfortunately this is not the case for a photon target with non-zero,
although small, virtuality $P^2$. The only measurement for the virtual
photon structure function available so far has been performed
by the PLUTO collaboration at PETRA \cite{8}. They measured the
structure function $F_{eff}=F_2+\frac{3}{2}F_L$ for $P^2 \le 0.8$
GeV$^2$ and $Q^2=5$ GeV$^2$ as a function of $x$ ($Q^2$ is the
virtuality of the probing virtual photon, whereas $P^2$ always denotes
the virtuality of the 
probed virtual target photon). More and better data should come from LEP2
\cite{9}. On the theoretical side several models exist 
for describing the $Q^2$-evolution equations of the parton distributions
and the input distributions at $Q_0$ with changing $P^2$ 
\cite{10,11, 12}. These constructions use  
essentially the same methods as have been applied for the parton
distributions of real photons. This way some smooth behavior
towards $P^2 = 0$, where previous results for the real photon should
hold, is guaranteed. In \cite{10} this construction allows a
calculation of the parton distribution functions (PDF) for virtual
photons in LO and NLO. It incorporates a purely perturbative
contribution and a non-pointlike hadronic contribution. Unfortunately
these PDF's for $P^2 \ne 0$ are not available in a form that parametrizes the
$Q^2$ evolution. Such parametrized PDF's for virtual photons have been
presented recently by two groups \cite{11, 12}, but unfortunately
only in LO. These two models differ somewhat in their method of
extrapolation to $P^2\ne 0$, in the choice of the input distribution and the
choice of the input scale $Q_0$. Furthermore the PDF's of Gl\"uck, Reya
and Stratmann (GRS) \cite{12} present distributions only for  $N_f= 3$
flavors, whereas Schuler and Sj\"ostrand (SaS) give PDF's for  $N_f= 4$
flavors \cite{11}. We expect these two structure function sets
to be detailed enough so they can be tested in photoproduction
experiments with virtual photons. First preliminary data have
been presented by the ZEUS \cite{13} and the H1 \cite{14}
collaborations. ZEUS studied the dijet cross section for $E_T > 4$ GeV
in the range of $0 < x_\g < 1$. Events with $x_\g > 0.75$ are
assumed to be dominated by the direct process whereas events with $0
< x_\g < 0.75$ give the resolved component. Then the ratio of the
direct-enriched and resolved-enriched cross section is measured as a
function of $P^2$ for $0<P^2<0.55$ GeV$^2$. In the H1 experiment the
inclusive one-jet cross section is measured as a function of $E_T$ and
rapidity $\eta$ for various $P^2$ bins \cite{14}. It is expected that
the resolved component decreases relative to the contribution from the
direct photon processes as the virtuality of the photon increases.
Some theoretical studies of the single inclusive and dijet inclusive
cross section in LO have been presented recently \cite{12, 17, 18}. 

It is well-known that in NLO calculations the distinction between
direct and resolved photoproduction becomes ambiguous. In this
order, a large contribution in the NLO direct cross section  is
subtracted from the direct component and combined with the LO
resolved term thus producing the scale ($Q\equiv M_\g$) 
dependence of the PDF's of the photon. Therefore both components are related
to each other through the factorization scale $M_\g$ at the photon leg
which determines the part of the NLO direct contributions to be
absorbed into the resolved component. The $M_\g$ dependence
of the remaining NLO direct contribution cancels to a large
extent against the  dependence in the resolved cross section
coming from the photon structure function. This connection has
been worked out in detail \cite{15, 4} and studied numerically
for real photoproduction with  $P^2= 0$ \cite{16}. It is clear that
this cancellation of the scale dependence must take place also
for $P^2\ne 0$. So, for a consistent calculation up to NLO one needs to
superimpose both contributions, the NLO direct and at least the LO
resolved cross section, both computed with the same  scale.
The subtraction of the large contribution in the direct cross section
which is shifted to the resolved term is defined only up to finite,
non-singular terms in the limit $P^2\to 0$. These finite terms may be
fixed in a way that a smooth behavior towards the limiting case
of real photons ($P^2= 0$) is achieved where these finite terms are
usually defined in the $\overline{\mbox{MS}}$ subtraction scheme. For
a complete NLO calculation we must include the NLO hard scattering
parton-parton cross sections for the direct (here one of the ingoing
partons is the photon with virtuality $P^2\ne 0$) and for the resolved process
together with the two-loop evolved structure function of the proton
and photon with virtuality $P^2\ne 0$.

In this paper we shall work out the subtraction in the NLO direct
contribution and superimpose the remaining direct contribution
and the resolved cross section up to NLO. We study this cross
section and the two contributions as a function of $P^2$ for various
inclusive one- and two-jet cross sections. Of particular interest
is the behavior of the resolved component as a function of $P^2$ and
the question at which $P^2$  the sum of direct and resolved cross sections
approaches the unsubtracted direct cross section which one expects
to give the correct description at sufficiently large $P^2$.

The outline is as follows. In section 2 we describe how to subtract the
singular terms in the NLO direct cross section connected with the
collinear singularity of the $\g\to q\bar{q}$ contribution. Here we
also define the finite terms which depend on the subtraction scheme
used for $P^2= 0$. Our result concerning the $P^2$ dependence of
various one- and two-jet cross sections are presented in section 3. In
this section we also compare with the preliminary data from ZEUS
\cite{13}. Section 4 contains a summary and an outlook for future
investigations.

%******************************************************************
%******************************************************************
\section{Subtraction of Photon Initial State Singularities}

The NLO corrections to the direct process become singular in the
limit $P^2= 0$. For real photoproduction these photon initial state
singularities are usually evaluated, i.e. regularized, with the
dimensional regularization method in which they are
finite for $\e =(4-d)/2$. The singular contributions appear as poles
in $\e$ multiplied by the splitting function $P_{q_i\leftarrow\g}$
\cite{19}. These initial state singularities are absorbed into the PDF
of the real photon ($P^2 = 0$). With no further finite terms subtracted
(for $\e\to 0$) this defines the $\overline{\mbox{MS}}$
factorization  scheme which must be consistently applied also for the
NLO evolution of the photon PDF. For $P^2\ne 0$ the corresponding
contributions appear as terms proportional to $\ln (P^2/s)$, $\sqrt{s}$
being the c.m.\ energy of the photon-parton subprocess. These terms are 
finite as long as $P^2\ne 0$ and can be calculated with $d = 4$ dimensions.
Since for small $P^2$ these terms are large they are absorbed as in the
case $P^2 = 0$ into the PDF of the virtual photon which is present in the
resolved cross section. Concerning finite terms (for $P^2\to 0$) which may
also be subtracted together with the singular terms we have the same
freedom as in the case $P^2 = 0$. We fix these finite terms such that
they agree with the $\overline{\mbox{MS}}$ factorization in the real
photon case. To achieve this we must compare the contributions
originating from the photon initial state singularities in the two
cases $P^2 = 0$ and $P^2\ne 0$, which we shall do in the following.

\begin{table}[bbb]
\renewcommand{\arraystretch}{1.6}
\caption{Classification of $2\to 3$ matrix elements}
\begin{center}
\begin{tabular}{|c|c|c|} \hline
 Class & Process & Class in \cite{19} \\ \hline\hline 
%------------------------------------------------
 $K_1$ & $\g q\to qgg$ & $I_1+I_2$ \\ \hline 
 $K_2$ & $\g g\to q\bar{q}g$ & $I_6+I_7$ \\ \hline 
 $K_3$ & $\g q\to qq\bar{q}$ & $I_3+I_4+I_5$ \\ \hline 
 $K_4$ & $\g q\to qq'\bar{q}'$ & $I_5$ \\ \hline 
\end{tabular}
\end{center}
\renewcommand{\arraystretch}{1}
\end{table}

For this purpose we must isolate the singular terms from the
photon initial state when the photon is collinear with one of
the outgoing quarks or antiquarks. The relevant processes are
labeled $K_1, K_2, K_3$ and $K_4$ as specified in Tab.\ 1. We do not
separate the contributions according to color factors as in
\cite{19}. To make the comparison with \cite{19} possible we have given
the contribution to the relevant processes also in the notation used
in \cite{19}. Using the same definitions of momenta and variables
as in \cite{19} the integration of the $2\to 3$ matrix elements  over
the phase space $d$PS$^{(r)}$ yields for $P^2 = 0$ the following result
($i = 1, 2, 3, 4$):
\equ{}{  \int d\mbox{PS}^{(r)} H_{K_i} = \int\limits_{x_a}^1
  \frac{dz_a}{z_a} e^2g^2\mu^{4\e}8(1-\e )\frac{\al_s}{2\pi}
  \left( \frac{4\pi\mu^2}{s} \right)^\e \frac{\G (1-\e )}{\G (1-2\e )}
    \frac{1}{4} K_i \quad . }
The result for the $K_i$ can easily be read from the results in
appendix C of \cite{19} by adding the corresponding sums of $I_i$'s
according to Tab.\ 1. The result is written in the following form
\begin{eqnarray}
  K_1 &=&  2C_F M U_1(s,t,u) \\
  K_2 &=&  -2M \bigg[ U_1(t,s,u) +U_1(u,s,t) \bigg] \\
  K_3 &=&  4C_FM \bigg[ U_s(s,t,u) + U_2(t,s,u) \nonumber \\ 
  &+& {\scriptstyle \frac{1}{2}} (U_3(s,t,u) + \mbox{cycl.\
  permutations in $s,t,u$}) \bigg] \\
  K_4 &=&  2C_FM \bigg[ U_3(s,t,u) + \mbox{cycl.\ permutations in
  $s,t,u$} \bigg] 
\end{eqnarray}
where $M$ is defined as
\equ{}{  M = -\frac{1}{\e} \frac{1}{2N_C} P_{q_i\leftarrow \g}(z_a)
  + \frac{1}{2N_C}P_{q_i\leftarrow \g}(z_a)\ln\left(
  \frac{(1-z_a)}{z_a} y_s\right) + \frac{Q_i^2}{2} \quad . }
In (6) $z_a=\frac{p_1p_2}{p_0q} \in [x_a,1]$ and the splitting function
\equ{}{ P_{q_i\leftarrow \g}(z_a) =
  2N_CQ_i^2 \ \frac{z_a^2+(1-z_a)^2}{2} \quad . }
The functions $U_i(s,t,u)$ are the LO parton-parton scattering cross
sections related to the various processes as shown in Tab.\ 2.
Processes, which are related by crossing are omitted, the complete
list is given in \cite{19}, $q$ and $q'$ denote quarks with different
flavors. The explicit expressions for the $U_i$ and their dependence on
color factors are given in \cite{19}. The factor $M$ contains the
characteristic singularity proportional to $1/\e$ which is
subtracted by absorbing
\equ{}{ R_{q_i\leftarrow\g}(z_a,M_\g^2) = 
  -\frac{1}{\e}P_{q_i\leftarrow\g}(z_a) \frac{\G (1-\e )}{\G (1-2\e )}
  \left( \frac{4\pi\mu^2}{M_\g^2} \right)^{\e} }
into the PDF of the photon $F_{a/\g}(x_a,M_\g^2)$ (see
\cite{19}). This subtraction produces the factorization scale
($M_\g$) dependence of the photon PDF and gives the finite
contributions to the cross section which are given by (2)--(5) with M
replaced by $M_{\overline{MS}}$: 
\equ{}{ M_{\overline{MS}} = -\frac{1}{2N_C} 
  P_{q_i\leftarrow\g}(z_a) \ln\left( \frac{M_\g^2z_a}{y_ss(1-z_a)}
  \right) + \frac{Q_i^2}{2} \quad .}
The variable $y_s$ which appears in (6) and (9) is the invariant mass
cut-off used to isolate the collinear singularity in the  
$\g\to q\bar{q}$ splitting.

\begin{table}[hhh]
\renewcommand{\arraystretch}{1.6}
\caption{LO parton-parton scattering matrix elements}
\begin{center}
\begin{tabular}{|c|c|c|} \hline
 Process & Matrix elements $|{\cal M}|^2 = 8N_CC_Fg^4U_i$ \\ \hline\hline
%----------------------------------------------------------
 $q\bar{q} \to gg$ & $U_1(s,t,u)$ \\ \hline 
 $qq'\to qq'$ & $U_2(s,t,u)$ \\ \hline 
 $qq\to qq$ & $U_2(s,t,u)+U_2(s,u,t)+U_3(s,t,u)$ \\ \hline 
\end{tabular}
\end{center}
\renewcommand{\arraystretch}{1}
\end{table}

The procedure for virtual photons with virtuality $P^2$ is completely
analogous. First one calculates the $2\to 3$ matrix elements, but now
with $P^2\ne 0$ and decomposes them into terms with the characteristic
denominator from $\g\to q\bar{q}$ splitting which become singular in
the limit $P^2\to 0$. These terms after phase space integration up to
a cut-off $y_s$ as in (1) have the same structure as (2)--(5) with the LO
parton-parton matrix elements factored out. The integration can be
done with $\e = 0$  since $P^2\ne 0$. The phase space in (1) then contains
the additional $P^2$-dependent factor $f$
\equ{}{ f = 1+ \frac{P^2(1-z_a)}{z_a(z_as-P^2)} }
which reduces to $1$ for $P^2=0$. The factor $M$
in the equations (2)--(5) takes the simple form
\equ{}{ M = \frac{1}{2N_C} \ln\left( 1 + \frac{y_ss}{z_aP^2}\right) 
  P_{q_i\leftarrow\g}(z_a) }
which is singular for $P^2 = 0$ as to be expected. This
singularity is absorbed into the PDF of the virtual photon with
virtuality $P^2$. Instead of (8) the subtraction term is:
\equ{}{ R_{q_i\leftarrow\g}(z_a,M_\g^2) = 
  \ln\left( \frac{M_\g^2}{P^2(1-z_a)} \right) P_{q_i\leftarrow\g}(z_a)
  -N_cQ_i^2 \quad . } 
After this subtraction the remaining finite term (for $P^2\to 0$) in $M$
yields 
\equ{}{ M(P^2)_{\overline{MS}} = -\frac{1}{2N_c}
  P_{q_i\leftarrow\g}(z_a) \ln\left(
  \frac{M_\g^2z_a}{(z_aP^2+y_ss)(1-z_a)} \right)  + \frac{Q_i^2}{2}
  \quad . } 
In addition to the singular term  $\ln (M_\g^2/P^2)$ we have
subtracted in (12) two finite terms in order to achieve in (13) the same
result as in (9) for $P^2 = 0$. Therefore we call this form of
factorization the $\overline{\mbox{MS}}$ factorization for 
$P^2\ne 0$. It is defined by the requirement that the remaining finite
term is equal to the finite term in (2)--(5) 
with $M$ replaced by $M_{\overline{MS}}$  as was obtained
in the calculation for real photons. With this definition of
factorization in the case  $P^2\ne 0$ we make sure that we obtain the same
NLO corrections as in \cite{19}, where $P^2=0$, when in the $P^2\ne 0$
calculation we choose $P^2$ extremely small. So, for the actual
calculations we apply the formulas (2)--(5) where $M$ is now given by
(13). We note that for averaging over the spin of initial gluons and
photons we apply a factor $1/(2-2\e )$, that gives rise to some additional
finite terms which should be included in (9) when expanded in $\e$. We
have taken these terms into account also in our calculation. This
completes the calculation of the contribution from the photon initial
state singularity.

It is clear that in NLO the PDF for the virtual photon must be
given also in the  $\overline{\mbox{MS}}$ factorization scheme. In
ref.\ \cite{10} the PDF is constructed in the so-called DIS$_\g$ scheme,
which is defined as for real photons ($P^2= 0$). This distribution
function is related to the  $\overline{\mbox{MS}}$ scheme PDF in the
following way \cite{10}: 
\equ{}{ F_{a/\g}(x,M_\g^2)_{DIS_\g} = 
  F_{a/\g}(x,M_\g^2)_{\overline{MS}} + 
  \de F_{a/\g}(x,M_\g^2) }
where for $a=q_i,\bar{q}_i$ or $g$:
\begin{eqnarray}
    \de F_{q_i/\g}(x,M_\g^2) &=&   \de F_{\bar{q}_i/\g}(x,M_\g^2)
    \nonumber \\
    &=& \frac{\al}{2\pi}N_C \bigg[ \frac{1}{N_C}
    P_{q_i\leftarrow\g}(x)   \ln\left( \frac{1-x}{x} \right) + Q_i^2
    8x(1-x) - Q_i^2 \bigg] \\
    \de F_{g/\g}(x,M_\g^2) &=& 0 \quad . \nonumber
\end{eqnarray}
If the PDF in this scheme is used to calculate the resolved cross
section one must transform the NLO finite terms in the direct
cross section. This produces a shift of
$M(P^2)_{\overline{MS}}$ as given in (13) by the same expression
as (15). Then the relation is: 
\equ{}{ M_{DIS_\g}(P^2)  = M(P^2) _{\overline{MS}}
  - N_C \bigg[ \frac{1}{N_C}P_{q\leftarrow \g}(z_a)\ln\left(
  \frac{1-z_a}{z_a} \right) 
  + Q_i^28z_a(1-z_a) - Q_i^2 \bigg] \quad . } 

All other singular terms in the real corrections, i.e.\ final
state singularities and the contributions from parton initial
state singularities have been calculated by Graudenz in connection
with the NLO corrections for jet production in deep inelastic
ep scattering. They can be taken from his work \cite{20} together
with the virtual corrections up to ${\cal O}(\al\al_s^2)$. (For related
work see \cite{21}.) All these contributions are calculated in dimensional
regularization. The appearing singularities in $1/\e$ cancel when all
singular terms are added. The remaining finite terms enter
the NLO corrections for the jet cross sections. These finite
terms depend on the phase space slicing parameter $y_s$ which is introduced
to separate the singular regions of final and initial state infrared
and collinear divergences.

%*******************************************************************
%*******************************************************************
\section{Inclusive One- and Two-Jet Cross Sections}

In this section, we present some characteristic numerical results for
one- and two-jet cross sections as a function of the virtuality $P^2$.
We consider the contributions of the direct and resolved components
and their sum. Partly we shall compare the NLO results with LO
predictions. The input for our calculation is the following. We
have chosen the CTEQ3M proton structure function \cite{22} which is
a NLO parametrization with $\overline{\mbox{MS}}$ factorization and 
$\Lambda^{(4)}_{\overline{MS}} = 239$ MeV. This $\Lambda$ value
is also used to calculate $\al_s$ from the two-loop formula at the scale $\mu
=E_T$. The factorization scales are also $M_\g=M_p=E_T$. We also need
the parton distribution of the virtual photon $F_{a/\g}$. For this we choose
either the GRS \cite{12} set or the SaS1M set of Schuler and
Sj\"ostrand \cite{11}. Both sets are given in parametrized form for
all scales $M_\g^2$ so that they can be applied without repeating the
computation of the evolution. Unfortunately both sets are given only
in LO, i.e. the boundary conditions for $P^2=M_\g^2$ and the evolution
equations are in LO. In \cite{10} PDF's for virtual photons have
been constructed in LO and NLO. Distinct differences occur for larger $P^2$
and  $x>10^{-3}$ which is mainly due to the different NLO perturbative boundary
condition at  $P^2=M_\g^2$, which does not exist for the real
($P^2=0$) photon structure function. Since neither of the two PDF's is
constrained by empirical data from  scattering on a virtual photon
target we consider these LO distribution functions as sufficient for our
exploratory studies on jet production and treat them as if they
were obtained in NLO. In particular, we shall use the transformation
formulae (14) for going from the DIS$_\g$  to the
$\overline{\mbox{MS}}$-scheme PDF, which makes sense only in NLO. Then
the parametrization \cite{12} is considered as the parametrization in
the DIS$_\g$ scheme. In the PDF of GRS the
input scale is $Q_0=0.5$ GeV and the restriction $P^2\le Q^2/5$ is
implemented as to fulfill the condition $\Lambda^2 \ll P^2 \ll
Q^2$. In addition the PDF of GRS is constructed only for $N_f=3$
flavors. The production of the heavier $c$ and $b$ quarks is supposed 
to be added as predicted by perturbation theory of fixed order with no
active $c$ and $b$ quarks in the proton and photon PDF's. In LO this
amounts to adding the processes $\g^*g\to c\bar{c}$ and $\g^*g\to
b\bar{b}$ to the cross section, keeping $m_c, m_b\ne 0$. Since in 
this work we are  primarily interested in studying the sum of the
direct and resolved contributions and the influence of the consistent
subtractions of the NLO direct part we refrain from adding the LO or
NLO cross sections for direct heavy quark production as suggested in
\cite{10, 12}. So, our investigations in connection with the GRS
parametrization of the virtual photon PDF are for a model with three
flavors only. For consistency we take also $N_f=3$ in the NLO
corrections and in the two-loop formula for $\alpha_s$. 
To overcome this problem we studied the relevant
cross sections also with the virtual photon PDF's of \cite{11}. Here
the $c$ quark is included as a massless flavor in the PDF which
undergoes the usual evolution as the other massless quarks except
for a shift of the starting scale $Q_0$. This $N_f=4$ PDF is considered
only in the parametrization SaS1M with  $Q_0= 0.6$ GeV that is in the
$\overline{\mbox{MS}}$ scheme.

The cross sections we have computed are essentially for kinematical
conditions as in the HERA experiments. There, positrons of $E_e = 27.5$ GeV
produce photons with virtuality $P^2$ which then collide with a proton
of $E_p = 820$ GeV. The momentum transfer to the photon is
$q=p_e-p_e'$ with  $P^2=(-q^2)$. The energy spectrum of the virtual
photons is approximated by 
\equ{ww}{ \frac{dF_{\g /e}(y)}{dP^2} =
  \frac{\al}{2\pi} \frac{1+(1-y)^2}{y} \frac{1}{P^2}  }
with  
\equ{}{ y = \frac{pq}{pp_e} \simeq \frac{E_{\g^*}}{E_e} }
being the fraction of the electron energy transferred to the photon,
when the virtuality $P^2\ll E_{\g^*}^2$. The momentum of the incoming proton
is $p$. The approximation for the virtual photon spectrum is used
for the calculation of the resolved and the direct cross section.
The expression (\ref{ww}) factorizes in the cross section for  $ep\to e'X$
with arbitrary final state $X$ if one neglects the longitudinal virtual
photon terms. After integration over $P^2$ between $P^2_{min} <
P^2<P^2_{max}$ with $P^2_{min}:=m_e^2y^2/(1-y)$, where $m_e$ is the
electron mass, one obtains the Weizs\"acker-Williams formula as used
for calculations with untagged electrons as in \cite{4, 19}, where
$P^2\simeq 0$ dominates. The cross section for the process $ep\to e'X$
is then given by the convolution
\equ{}{ \frac{d\si (ep\to e'X)}{dP^2} = \int\limits_{y_{min}}^{y_{max}}
    dy \frac{dF_{\g /e}(y)}{dP^2} d\si (\g^*p\to X) } 
where $d\si (\g^*p\to X)$ denotes the cross section for  $\g^*p\to
X$ with transversely polarized photons of energy
$q_0=E_ey-P^2/(4E_ey)$, if the transverse component of $q$ is neglected. 
To have the equivalent conditions as in the ZEUS analysis we choose
$y_{min}= 0.2$ and $y_{max}=0.8$. In the computation of the resolved
cross section the approximation $q_0=E_ey$ is inserted for the energy of the
virtual photon, whereas for the direct photon cross section the exact
relation for $q_0$ is taken into account through the kinematic relations. 

All further calculations proceed in the following way. For both,
the direct and the resolved cross section, we have a set of two-body
contributions and a set of three-body contributions. Each set is
completely finite, as all singularities have been canceled or
absorbed into PDF's. Each part depends separately on the phase space
slicing parameter $y_s$. The analytic calculations are valid only for
very small $y_s$, since terms ${\cal O}(y_s)$ have been neglected in
the analytic integrations. For very small $y_s$, the two separate
pieces have no physical meaning. In this case the $(\ln y_s)$-terms
force the two-body contributions to become negative, whereas the
three-body cross sections are large and positive. In \cite{3} we have
plotted such a cross sections for direct real photoproduction at $y_s =
10^{-3}$. When the two-body and three-body contributions are
superimposed to yield a suitable inclusive cross section, as for
example the inclusive one- or two-jet cross section, the dependence on
the cut-off $y_s$ will cancel. The separation of the two contributions
with the slicing parameter $y_s$ is a purely technical device in order to
distinguish the phase space regions, where the integrations are done
analytically, from those, where they are done numerically. We have
checked, by varying $y_s$ between $10^{-4}$ and $10^{-3}$, that the
superimposed two- and three-body contributions are independent of $y_s$
for the inclusive single- and dijet cross sections.

First, we study the inclusive single-jet cross section. To calculate
it we must choose a jet definition, which combines two nearly
collinear partons. We adopt the jet definition of the Snowmass
meeting \cite{23}. According to this definition, two partons $i$ and $j$
are recombined, if $R_{i,j}<R$, where 
$R_{i,J}=\sqrt{(\eta_i-\eta_J)^2+(\phi_i-\phi_J)^2}$ and  $\eta_J, \phi_J$
are the rapidity and the azimuthal angle of the combined jet
respectively, defined as
\begin{eqnarray}
 E_{T_J} &=& E_{T_1} + E_{T_2} \\
 \eta_J  &=& \frac{E_{T_1}\eta_1 + E_{T_2}\eta_2}{E_{T_J}} \quad , \\
 \phi_J  &=& \frac{E_{T_1}\phi_1 + E_{T_2}\phi_2}{E_{T_J}} \quad , 
\end{eqnarray}
and $R$ is chosen as in the experimental analysis. So, two partons
are considered as two separate  
jets or as a single jet depending whether they lie outside or inside
the cone with radius $R$ around the jet momentum. In NLO, the final state
may consist of two or three jets. The three-jet sample contains
all three-body contributions, which do not fulfill the
cone condition. The cone constraint is evaluated in the HERA
laboratory system as for real photoproduction ($P^2 = 0$) and in the
experimental analysis. The calculation of the resolved cross section
proceeds as for real photoproduction, i.e. the transverse momentum
($q_T$) of the virtual photon is neglected so that the virtual photon momentum
is in the direction of the incoming electron and $q_0=E_ey$. The cross
section for direct photons, in which the initial state singularity is
subtracted, as specified in the previous section, is given in the
center-of-mass system ${\bf p}+{\bf q}={\bf 0}$ and transformed into
the HERA laboratory system, again neglecting  $q_T$ and other small terms
proportional to $P^2$. Then the relation between the rapidity
$\eta_{cm}$ of the jet in the c.m.-system and laboratory system is as
for real photoproduction, which is
\equ{}{ \eta = \eta_{cm} + \frac{1}{2} \ln\frac{E_p}{yE_e} \quad . }
This $\eta$ and the corresponding azimuthal angle of the partons in the
final state is also used for evaluating the jet definition given
above.

In Fig.\ 1a, b, c, the results for $d^3\si /dE_Td\eta dP^2$ are shown
integrated over $\eta$ in the interval $-1.125 \le \eta \le 1.875$
(these boundaries are employed in the ZEUS analysis \cite{13}) as a
function of $E_T$ for the three values of $P^2 =0.058,0.5$ and $1$
GeV$^2$. The smallest value of $P^2$ has been chosen in such a way
that it reproduces the $P^2 \simeq 0$ results. Inserting $P^2=0.058$
GeV$^2$ into the unintegrated Weizs\"acker-Williams formula
corresponds to the one integrated in the region $P^2_{min} \le P^2 \le
P^2_{max} = 4$ GeV$^2$. For all three $P^2$ the cross section is
dominated by the resolved component at small $E_T$. Near $E_T=20$ GeV
the direct component, which is the direct cross section without the
subtraction term, denoted by Dir$_s$, is of the same magnitude as the
resolved cross section. As a function of $P^2$ the cross section 
$d^3\si /dE_Td\eta dP^2$ decreases more or less uniformly in the
considered $E_T$ range with increasing $P^2$. One can see that this
decrease is stronger for small $E_T$ as compared to large $E_T$. In
these and the following results the cone radius is $R=1$. The
corresponding rapidity distribution for a fixed $E_T$ will be shown
later in Fig.\ 5a, b, c, d for the SaS1M photon PDF. 

In Fig.\ 2a, b, c we studied the $\eta$ distribution of the Dir$_s$
contribution at fixed $E_T=7$ GeV and the same $P^2$ values as in
Fig.\ 1 in comparison with two approximations, namely the LO cross
section and the NLO cross section with $P^2\ne 0$ only in the photon
flux equation (17) and the rest of the cross section evaluated at
$P^2=0$ as in \cite{19}. As to be expected this approximation is very
good for $P^2=0.058$ GeV$^2$. At the larger $P^2$ however it
overestimates the cross section and should not be used. This means that
the $P^2$ dependence of the direct part, although the strongest
logarithmic $P^2$ dependence has been subtracted, should be taken into
account. Of course, in the sum of the resolved and the direct cross
section the difference is small as long as the resolved part dominates
which is true for the smaller $E_T$'s. The LO prediction, which is
evaluated with the same structure functions and $\al_s$ value as the
NLO result, only the hard scattering cross section is calculated in
LO, is smaller than the NLO result, the difference decreasing with
increasing $P^2$. Of course, this finding depends on the chosen value
of $R$. The NLO cross section depends on $R$, whereas the LO curve
does not. Therefore estimates of the inclusive cross section with LO
calculations can be trusted only for large cone radii.

It is clear that the resolved and the direct cross sections decrease
with increasing $P^2$ for fixed $\eta$ and $E_T$. It is of interest to
know how the ratio of resolved and the direct cross section behaves
as a function of $P^2$. This is a well defined quantity in LO. The
variation of this ratio with $P^2$ up to $P^2=4$ GeV$^2$ for $E_T=7$
GeV and $\eta=2, 1, 0$ and $-1$ is plotted in Fig.\ 3a, b, c, d. As we
expected this ratio decreases most strongly for $\eta=2$, since in the 
$\eta >0$ region the resolved component dominates whereas the direct
cross section peaks for negative $\eta$'s (see Fig.\ 2a, b, c). If we
decrease $\eta$ towards the backward direction this ratio diminishes
more or less uniformly for all $P^2$. In NLO this ratio depends on
the scheme chosen for the photon PDF. In the DIS$_\gamma$ subtraction
scheme terms in the photon PDF are shifted to the direct cross section
as follows from (14)--(16) in section 2. This necessarily changes the
ratio of the resolved to the direct cross section, not only in the
absolute value but also in the dependence on $P^2$. For all $\eta$'s
this ratio is quite different from its LO value. The difference
between the $\overline{\mbox{MS}}$ and DIS$_\g$ scheme is small for
$\eta =0,1$ moderate for $\eta =2$ and of the order of $1.5$ at 
$\eta =-1$. Except at $\eta =-1$ the ratio is always larger in the
DIS$_\g$ scheme than in the $\overline{\mbox{MS}}$ scheme. Apart from
the fact that the ratios plotted in Fig.\ 3a, b, c, d cannot be measured
directly, they are scheme dependent in NLO and have very large
corrections when going from LO to NLO. In Fig.\ 3a, b, c we have also plotted
the ratio for the case that the $P^2$ dependence is neglected in
the cross section $\g^*p\to X$ and taken into account only in the
photon propagator (denoted NLO($P^2=0$)). This approximation gives a
result very similar to the LO curve showing that the NLO corrections
are more important for $P^2\ne 0$. Of course at $P^2\to 0$ this
approximation is equal to the NLO result in the $\overline{\mbox{MS}}$
scheme. 

The results shown so far are for a model with three flavors only and
therefore should not be compared to the experimental data except when
the contribution from the charm quark is added at least in LO. A more
realistic approach is to use the photon PDF's SaS1M of ref.\ \cite{11}
which are constructed for four flavors. The results for 
$d^3\si /dE_Td\eta dP^2$ integrated over $\eta\in [-1.125,1.875]$ as a
function of $E_T$ for $P^2=0.058, 0.5$ and $1.0$ GeV$^2$ are presented
in Fig.\ 4a, b, c. These curves can be compared with the results in
Fig.\ 1a, b, c obtained with the GRS photon distribution with
$N_f=3$. The results for the sum of the resolved and direct
contributions change between 10\% and 30\% in the small $E_T$ region
and approximately 50\% in the large $E_T$ region. As one can 
see the larger cross section for $N_f=4$ results primarily from the
direct component. Since the direct component is more important for
large $E_T$ than for small $E_T$ the increase is stronger in the
former region. 

Of interest are also the rapidity distributions for fixed $E_T$. These
are shown for $E_T=7$ GeV as a function of $\eta$ between $-1\le \eta
\le 2$ choosing $P^2=0.058,1,5$ and $9$ GeV$^2$ in Fig.\ 5a, b, c, d. We
show the subtracted direct cross section and the resolved cross section
and their sum for the photon PDF as in Fig.\ 4a, b, c. Expectedly the
resolved component has its maximum shifted to positive $\eta$'s as
compared to the direct component. The direct component falls off with
increasing $\eta$ quite strongly. This comes from the subtraction of
the $(\ln P^2)$ terms as is apparent when we compare with the
unsubtracted direct cross section, labeled "Dir" in Fig.\
5a, b, c, d. The sum of the resolved and the direct (subtracted) cross
section is more or less constant for the smaller $P^2$ values and
decreases with increasing $\eta$ for $P^2=5$ and $9$ GeV$^2$. 

For large enough $P^2$ we expect the unsubtracted direct cross section
to be the correct one. In this region the subtraction term (12) must
approximate the PDF of the photon rather well. We have checked this by
a direct comparison. With increasing $P^2$ the full direct cross
section (Dir) approaches the Sum=Res+Dir$_s$ better and better. But
even at $P^2=9$ GeV$^2$ the two cross sections differ at $\eta =2$
still by approximately 30 \%. However, in the backward direction at
$\eta =-1$ we also see a difference. In this region, which corresponds
in the photon PDF to the region $x_\g \simeq 1$, where the perturbative
component dominates, we do not expect a deviation. This may be caused
by our approximation of neglecting the transverse momentum $q_T$ of
the virtual photon in the relation (22) and in the calculation of the
resolved cross section. 

In order to compare with preliminary data for the dijet cross section
presented by the ZEUS collaboration \cite{13} we calculated the
inclusive dijet cross section $d^4\si/dE_Td\eta_1d\eta_2dP^2$ as a
function of $P^2$. Here $E_T$ is the transverse energy of the measured
or trigger jet with rapidity $\eta_1$. $\eta_2$ denotes the rapidity
of another jet such that in the three-jet sample these two jets have
the largest $E_T$, i.e.\ $E_{T_1}, E_{T_2} > E_{T_3}$. The
calculational procedure is analogous to the inclusive single-jet cross
section and follows closely the work presented in \cite{4} for
$P^2=0$. Since inclusive two-jet cross sections depend on one more
variable they constitute a much more stringent test of QCD predictions
than inclusive one-jet cross sections. We could predict distributions
in $\eta_1$ and $\eta_2$ for fixed $E_T$ or distributions in $E_T$ for
various values of the two rapidities $\eta_1$ and $\eta_2$ in the same
way as in \cite{4} for $P^2=0$. Since no such information is expected
from the experiment in the near future we calculated only the $E_T$
distribution with the two rapidities integrated over the interval 
$-1.125 < \eta_1,\eta_2 < 1.875$ following the constraints in the ZEUS
analysis \cite{13}. The results for $P^2=0.058, 0.5$ and $1.0$ GeV$^2$
are shown in Fig.\ 6a, b, c, where the full curve is 
$d^4\si/dE_Td\eta_1d\eta_2dP^2$ as a function of $E_T$ integrated over
$\eta_1$ and $\eta_2$ in the specified interval and for
$0.2<y<0.8$. The functional dependence on $E_T$ does not change as a
function of $P^2$, only the absolute value of the cross section
decreases with increasing $P^2$.

In Fig.\ 6a, b, c we show also the cross sections for enriched
direct and resolved $\gamma$ samples. These cross sections are labeled
"Dir" (dashed curve) and "Res" (dotted curve), respectively. The two
cross sections are defined with a cut on the variable $x_\gamma^{obs}$
where $x_\gamma^{obs}$ is defined by
\equ{}{ x_\gamma^{obs} = \frac{\sum_i E_{T_i}e^{-\eta_i}}{2yE_e}
  \quad . }
This variable measures the fraction of the proton energy that goes
into the production of the two hardest jets. The "Dir" curve gives
the cross section for $x_\gamma^{obs}>0.75$. This cut on
$x_\gamma^{obs}$ leads to an enrichment of the direct component of
the cross section, since exclusive two-jet events from the direct
process have $x_\gamma^{obs}=1$. The curve labeled "Res" gives the
cross section in the complementary region $x_\gamma^{obs}<0.75$ which
characterizes the enriched resolved $\gamma$ sample. The sum of the
Dir and Res curves is equal to the full cross section 
$d^4\si/dE_Td\eta_1d\eta_2dP^2$ with no cut on $x_\gamma^{obs}$. 
It must be emphasized that both cross sections, whether
$x_\g^{obs}>0.75$ or $x_\g^{obs}<0.75$, contain contributions from the
direct and resolved part. Actually the resolved contribution in the
enriched direct sample ($x_\g^{obs}>0.75$) is non-negligible. 
The results in Fig.\ 6a, b, c are for the GRS
parton distributions in the $\overline{\mbox{MS}}$ scheme. As to be
expected, with increasing $P^2$ the full cross section is more and
more dominated by the Dir component, in particular at the larger
$E_T$. This means that the cross section in $x_\gamma^{obs}<0.75$
decreases stronger with $P^2$ than in the $x_\gamma^{obs}>0.75$
region. This could be studied experimentally by measuring the ratio of
the two cross sections as a function of $P^2$ for fixed $E_T$. This
has not been done yet. Instead, the ZEUS collaboration \cite{13}
measured the ratio $r=$Res$/$Dir, where Res and Dir are
the cross sections as defined above, but integrated over
$E_{T_1},E_{T_2}\ge 4$ GeV. With this integration cut on the
transverse momenta of the two hardest jets, the transverse momentum of
the third jet may vanish which affords particular constraints on the
remnant jets. This is treated as in \cite{3}. Furthermore we replaced
the GRS photon PDF by the SaS1M photon PDF which is for $N_f=4$
flavors. With these assumptions we calculated the ratio $r$ as a
function of $P^2$ up to $P^2=0.6$ GeV$^2$ and compared it with the
ZEUS \cite{13} data in Fig.\ 7 in LO (dotted curve) and NLO (full
curve). The theoretical NLO curve agrees quite  
well with the data at $P^2 \ge 0.25$ GeV$^2$ but not with the measured
point at $P^2 \simeq 0.2$ GeV$^2$ and at $P^2 = 0.058$ GeV$^2$
corresponding to the photoproduction case. Of course the ratio $r$
for $P^2\simeq 0$ is much more precise and lies 30 \% higher than the
predicted cross section. This disagreement at $P^2\simeq 0$ is to be
expected since at this value of $P^2$ the measured inclusive dijet cross
section for the enriched resolved $\gamma$ sample is larger than the
predicted cross section for a small $E_T$ cut \cite{4}. As in \cite{4}
we attribute this difference between theory and experimental data to
additional contributions due to multiple interactions with the proton
remnant jet in the resolved cross section not accounted for by our NLO
predictions. This underlying event contribution is reduced with
increasing $E_T^{min}$ and/or smaller cone radii $R<1$. As it seems,
for larger $P^2$, the underlying event contribution is also
reduced. Of course this could be studied more directly by measuring
rapidity distributions for the enriched resolved $\g$ sample as was
done at $P^2\simeq 0$ in \cite{4}.

%********************************************************
\section{Conclusions}
We have calculated inclusive single jet and dijet cross sections for
photoproduction with virtual photons. The direct and resolved contributions
were calculated in next-to-leading order QCD using the phase-space slicing
method. They were folded with the unintegrated Weizs\"acker-Williams
approximation
and existing LO parametrizations for the virtual photon parton densities.
The collinear singularity in the direct photon initial state
was integrated analytically up to an invariant mass cut-off $y_s$. Contrary
to real photons, this specific singularity is
not regulated in the dimensional regularization scheme but
by the mass of the virtual photon $P^2$ leading to a logarithmic
dependence on $P^2$. This logarithmic term is absorbed into the
virtual photon structure 
function rendering the latter scheme and scale dependent. The remaining
finite contribution is constructed in such a way that the corresponding
real photon term is obtained in the
limit $P^2=0$ in the $\overline{\rm MS}$ scheme. Similarly to the 
construction of virtual photon structure functions by GRS and SaS, our
calculation of the hard scattering cross section then provides a consistent
extension from $P^2=0$ to small, but non-zero $P^2$.

We presented $y_s$-cut independent results using the Snowmass jet definition
and HERA conditions for distributions in the transverse energy and the
rapidity of the observed jet and in the photon virtuality. For the
case of very small $P^2$, we found good numerical agreement
with the predictions for real photoproduction.
At $P^2=9$ GeV$^2$, the unsubtracted direct contribution corresponding
to the case of deep inelastic scattering approximates the sum of the
subtracted direct and resolved contributions quite well. A small
discrepancy remains in the forward region, where the resolved
contribution is dominant. As in
the case of $P^2=0$, the resolved component dominates at low $E_T$ and in the
forward region. The NLO effects were demonstrated to be important in the
ratio of resolved and direct cross sections as a function of $P^2$. Since the
theoretical separation between direct and resolved is artificial, some
scheme dependence remains here unless both contributions are defined
by an experimental cut on $x_{\gamma}^{obs}$ in dijet cross sections.
Then the corresponding ratio shows significant NLO effects and good agreement
with the available ZEUS data for $P^2>0.2$ GeV$^2$. The disagreement below
this value can be attributed to additional contributions coming from, e.g.,
multiple scattering between the photon and proton remnants.

Future investigations on virtual photoproduction will require more data on
single inclusive jet production as they exist for $P^2=0$ and at
larger transverse energies. With luminosity permitting, a detailed dijet
analysis of an infrared safe cross section such as $d^4\sigma/dE_Td\eta_1
d\eta_2dP^2$, where the transverse energies of the two jets are not cut
at exactly the same value, will provide a much improved insight into the
structure of the virtual photon. Furthermore, choosing a $k_T$-cluster-like
jet definition with smaller cone radii will reduce both the uncertainties
in the jet algorithm and in the underlying event. On the theoretical side,
one possible improvement is the correct treatment of the transverse momentum
of the incoming photon for larger $P^2$ including a correct
transformation from the photonic c.m.~frame to the HERA laboratory
system. For a consistent NLO treatment, the inclusion of NLO parton
densities for the photon is necessary. These are, however, needed in a
parametrized form and should also be studied in correlation with deep
inelastic $e\gamma^\ast$ scattering data.

%*********************************************************************
%*********************************************************************

\newpage
%*********************************************************************
% 1) 1-jet inclusive pt-spectrum, eta=[-1.125,1.875] (GRS)
%*********************************************************************

\begin{figure}[hhh]
  \unitlength1mm
  \begin{picture}(122,150)
    \put(-2,-20){\epsfig{file=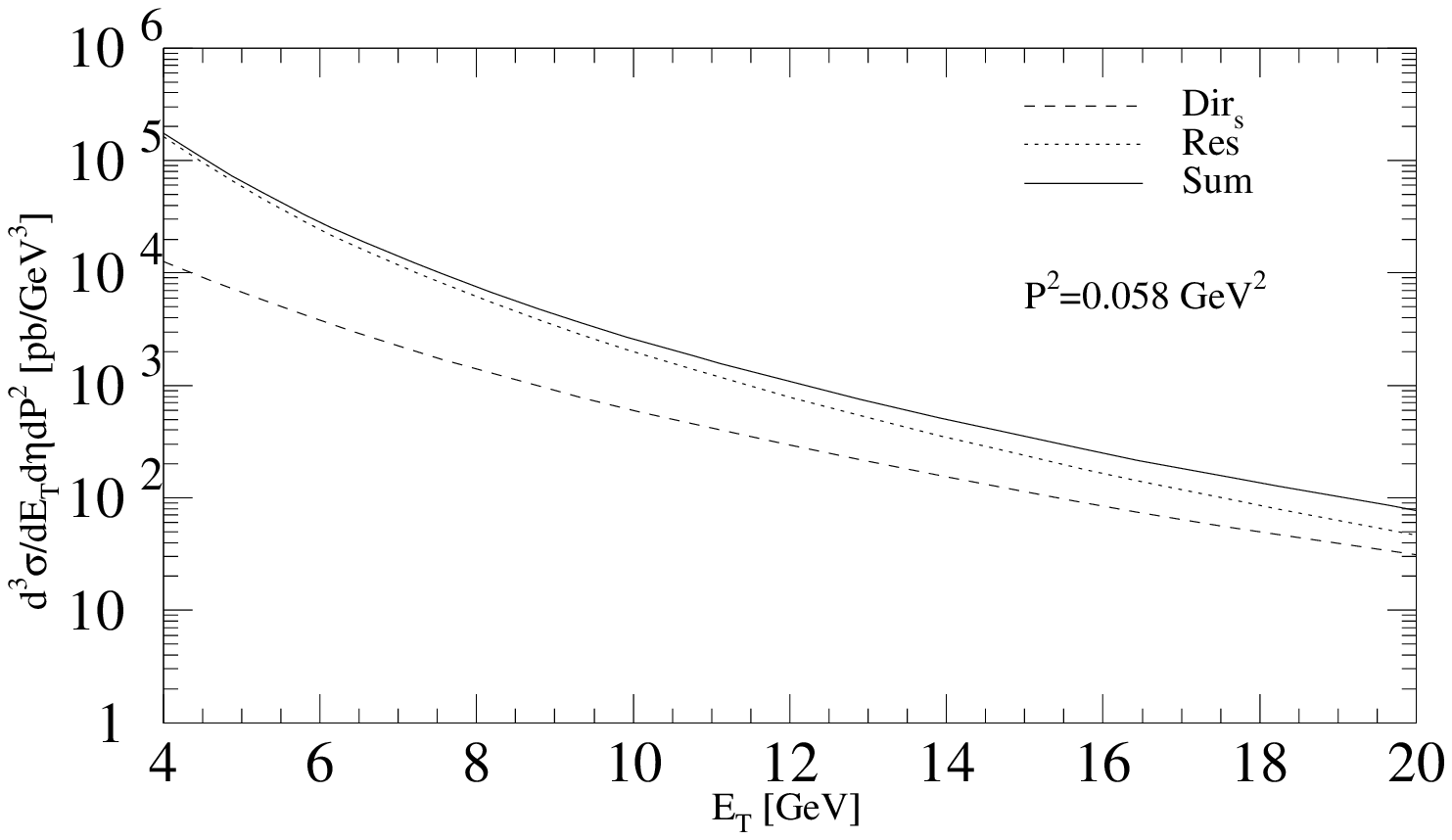,width=17cm}}
    \put(-2,-140){\epsfig{file=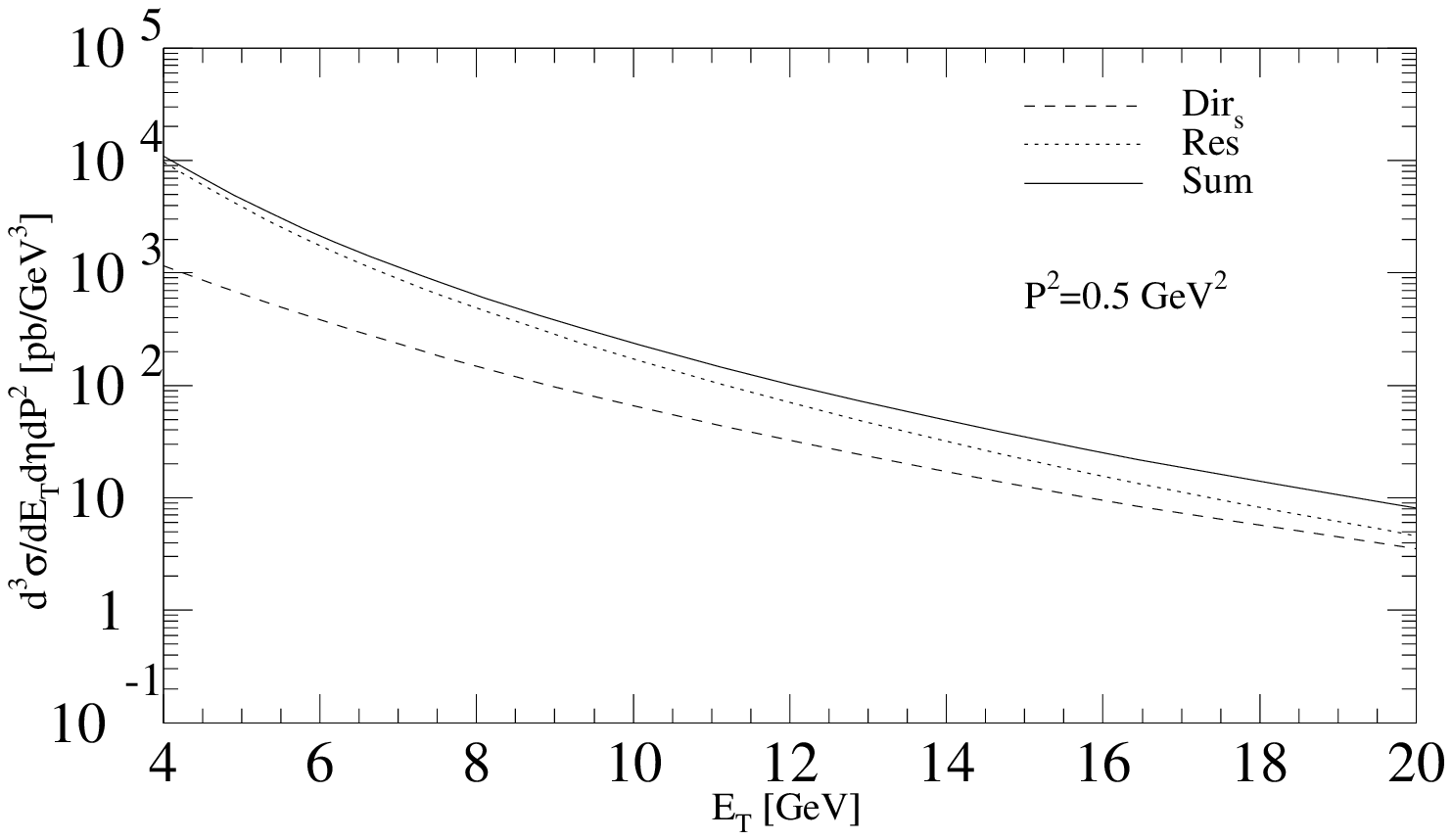,width=17cm}}
    \put(0,66){\parbox[t]{16cm}{\sloppy Figure 1a: Single-jet
        inclusive cross section integrated over $\eta \in
        [-1.125,1.875]$ for the virtuality $P^2=0.058$ GeV$^2$. The
        $\overline{\mbox{MS}}$-GRS
        parametrization with $N_f=3$ is chosen. The solid line gives
        the sum of the subtracted direct and the resolved term.}} 
    \put(0,-54){\parbox[t]{16cm}{\sloppy Figure 1b: Same as figure 1a
        with $P^2=0.5$ GeV$^2$.}}
  \end{picture}
\end{figure}
\newpage

\begin{figure}[hhh]
  \unitlength1mm
  \begin{picture}(122,150)
    \put(-2,-20){\epsfig{file=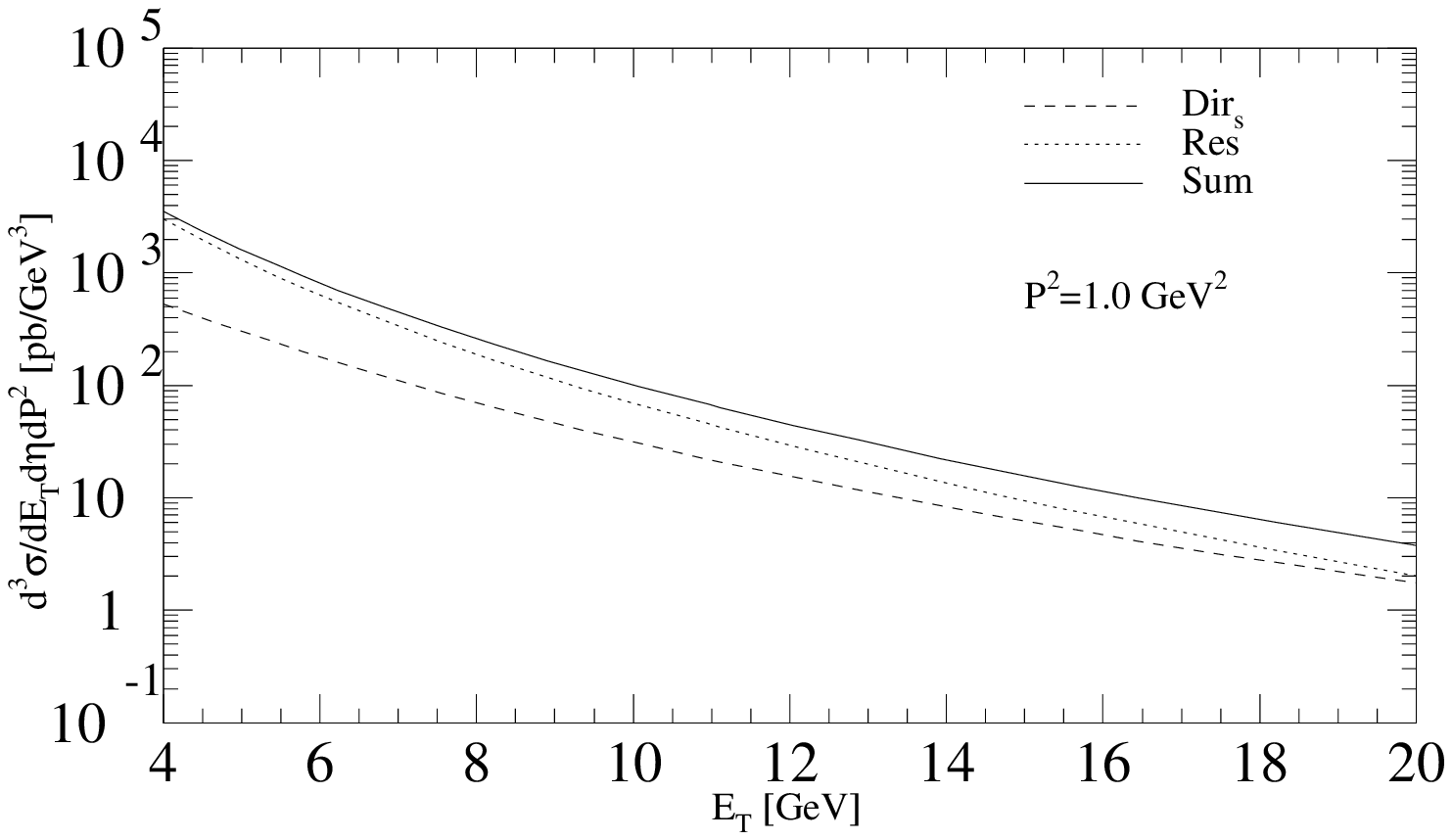,width=17cm}}
    \put(0,66){\parbox[t]{16cm}{\sloppy Figure 1c: Same as figure 1a
        with $P^2=1.0$ GeV$^2$.}}
  \end{picture}
\end{figure}
\newpage

%*********************************************************************
% 2) 1-jet inclusive eta-spectrum comparison NLO poeter/klasen (GRS)
%*********************************************************************

\begin{figure}[hhh]
  \unitlength1mm
  \begin{picture}(122,150)
    \put(-2,-20){\epsfig{file=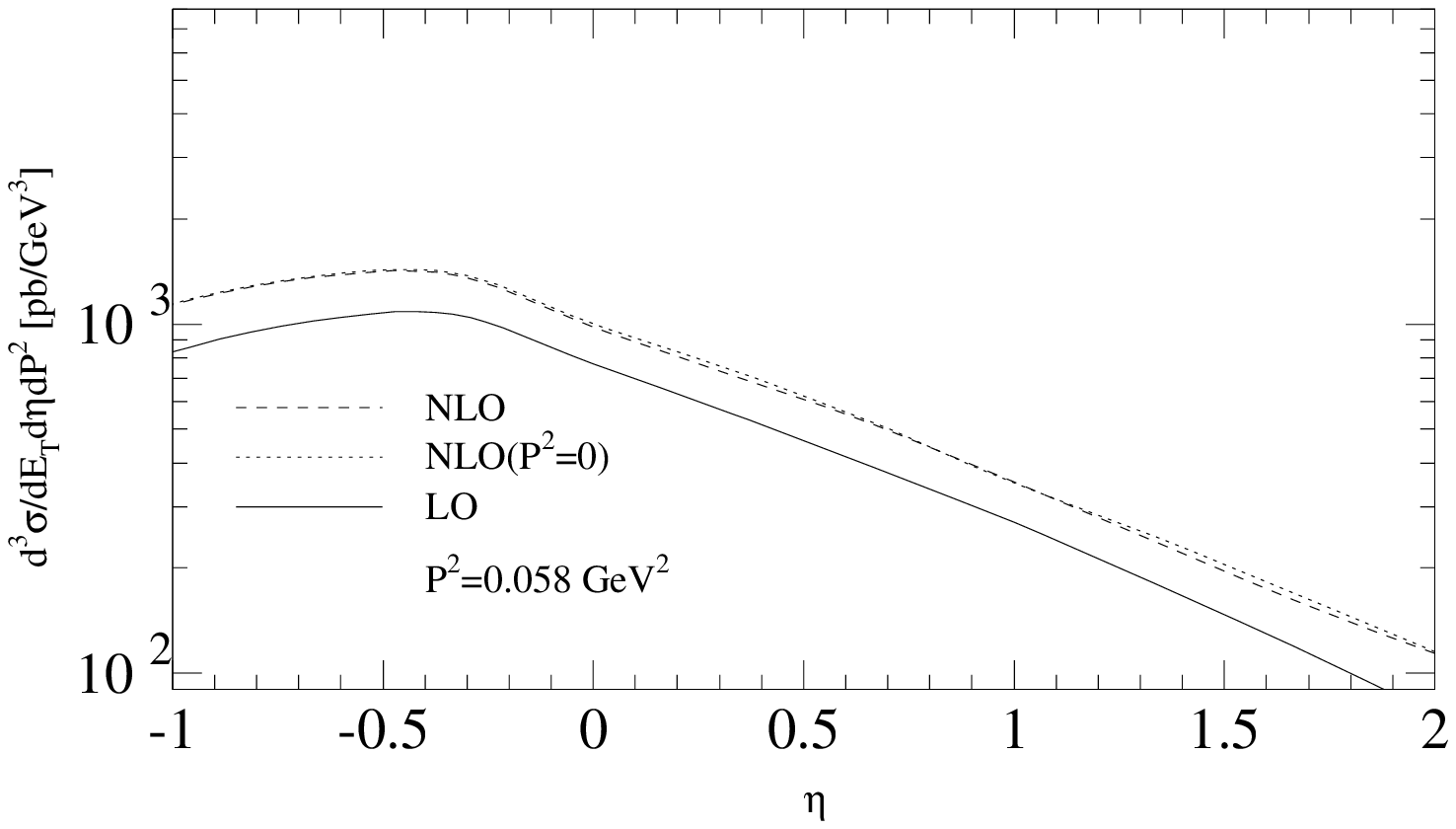,width=17cm}}
    \put(-2,-140){\epsfig{file=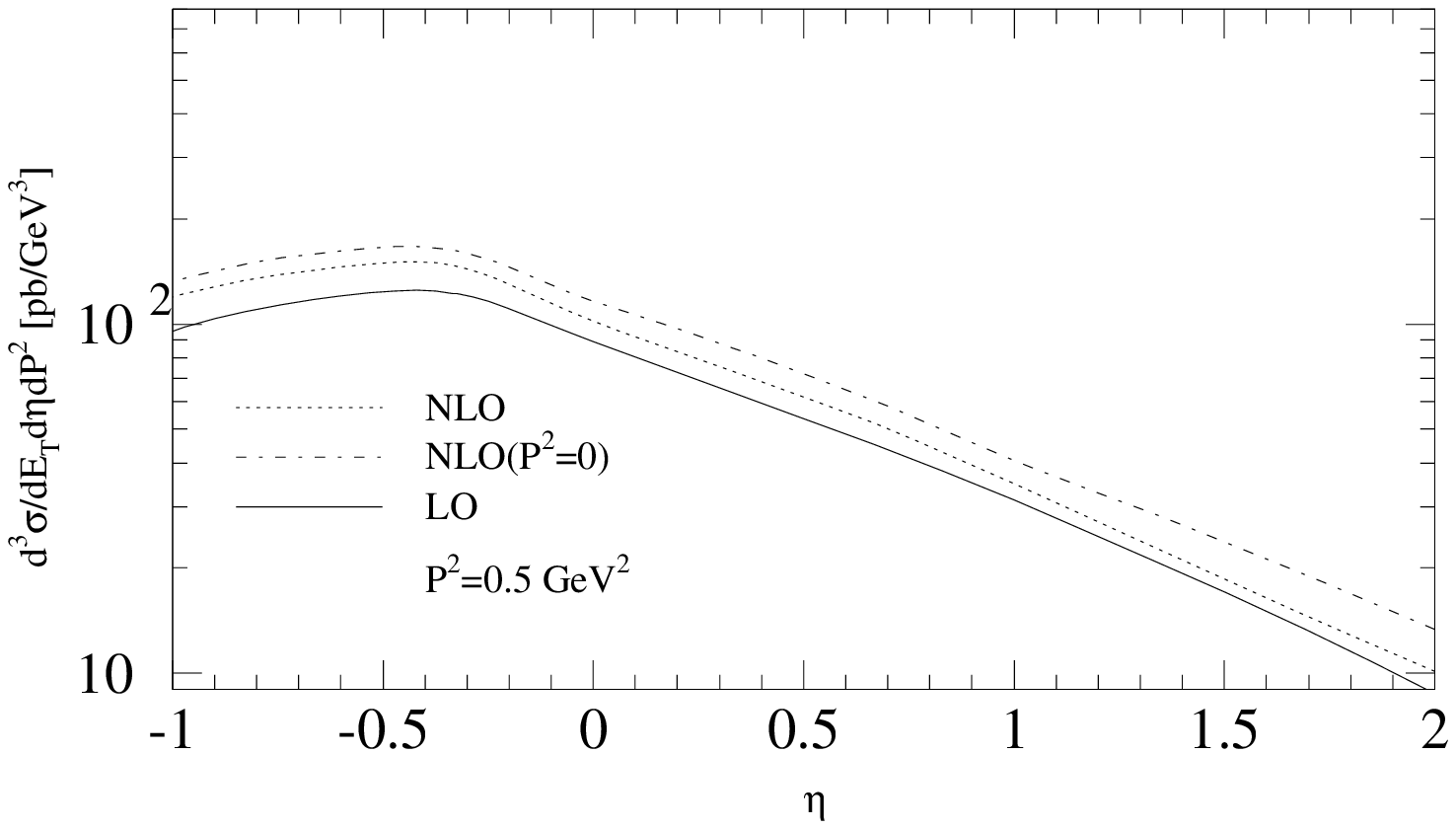,width=17cm}}
    \put(0,66){\parbox[t]{16cm}{\sloppy Figure 2a: Single-jet
        inclusive cross sections for $E_T=7$ GeV and $P^2=0.058$
        GeV$^2$. The $\overline{\mbox{MS}}$-GRS parametrization with
        $N_f=3$ is chosen. Only the direct part with subtraction
        (Dir$_s$) is  plotted. The solid line gives the LO
        contribution. The dashed curve is the full NLO cross section,
        whereas the dotted curve gives the NLO cross section,
        where the NLO matrix elements have no $P^2$-dependence.}}
    \put(0,-54){\parbox[t]{16cm}{\sloppy Figure 2b: Same as figure 2a
        with $P^2=0.5$ GeV$^2$.}}
  \end{picture}
\end{figure}
\newpage

\begin{figure}[hhh]
  \unitlength1mm
  \begin{picture}(122,150)
    \put(-2,-20){\epsfig{file=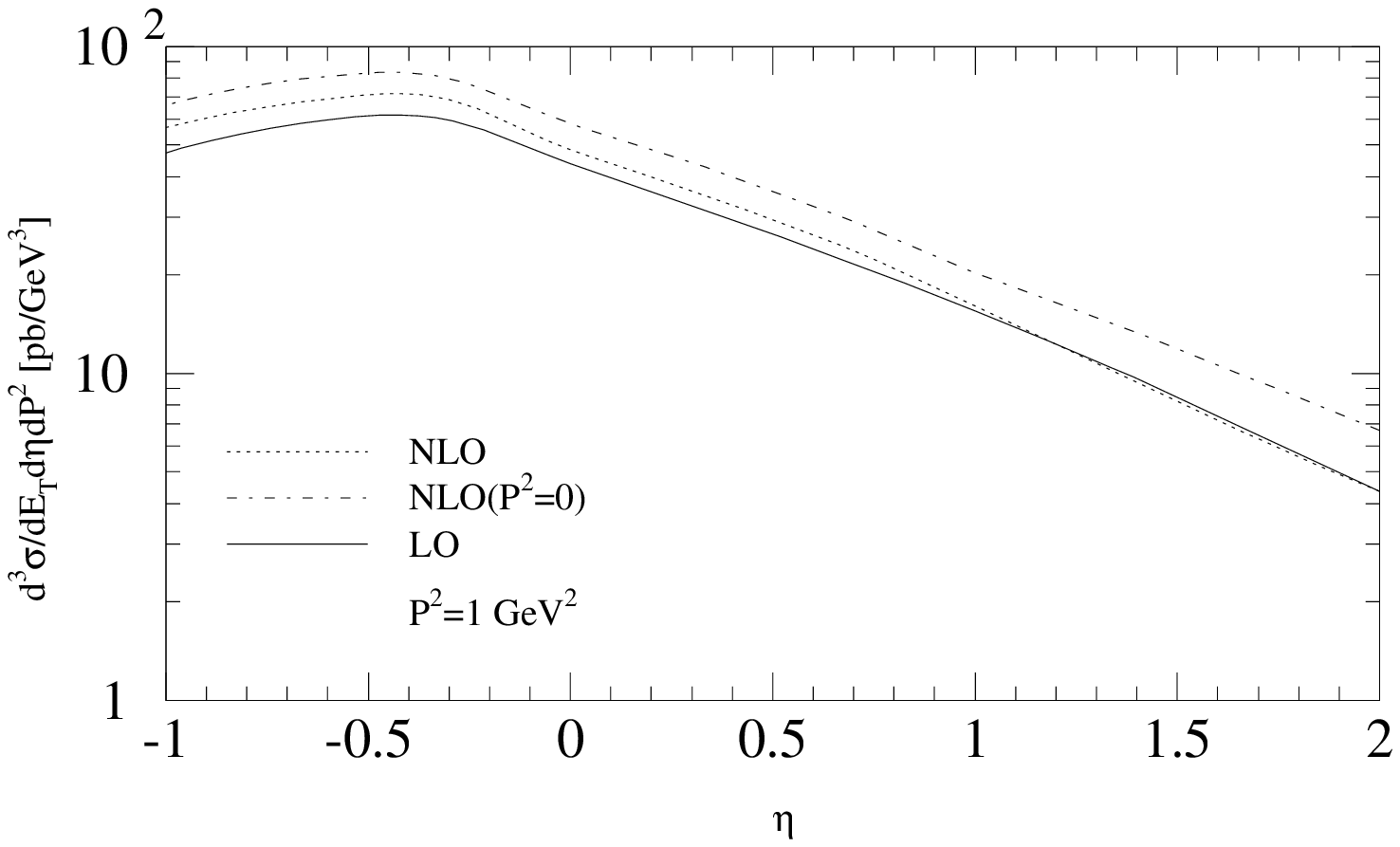,width=17cm}}
    \put(0,66){\parbox[t]{16cm}{\sloppy Figure 2c: Same as figure 2a
        with $P^2=1.0$ GeV$^2$.}}
  \end{picture}
\end{figure}
\newpage

%*********************************************************************
% 3) Ratio Res/Dir as P^2-spectrum, eta={2,1,0,-1}, pt=7 GeV (GRS)
%    MS-bar and DIS-g schemes.                                
%*********************************************************************

\begin{figure}[hhh]
  \unitlength1mm
  \begin{picture}(122,150)
    \put(-2,-20){\epsfig{file=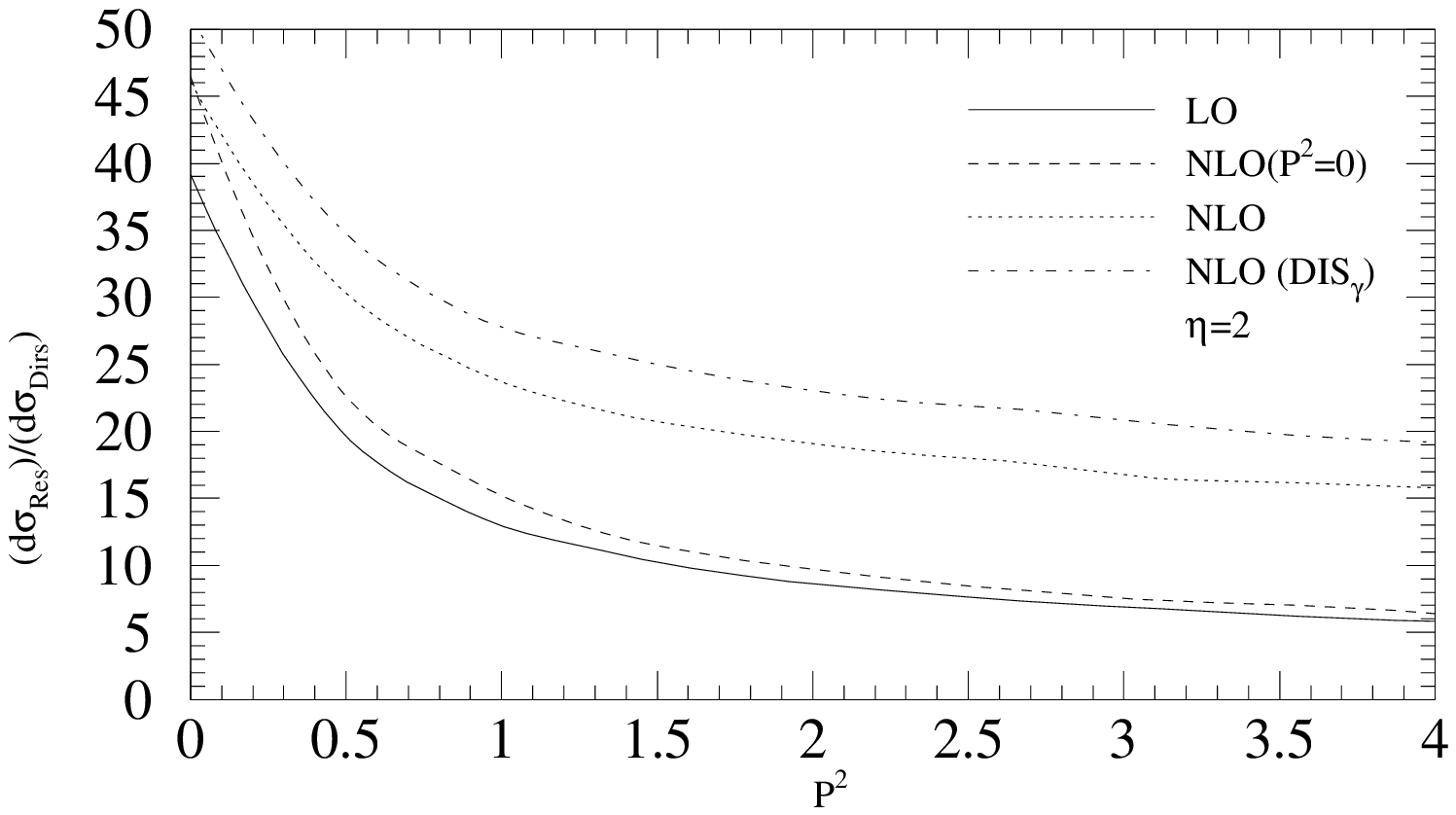,width=17cm}}
    \put(-2,-140){\epsfig{file=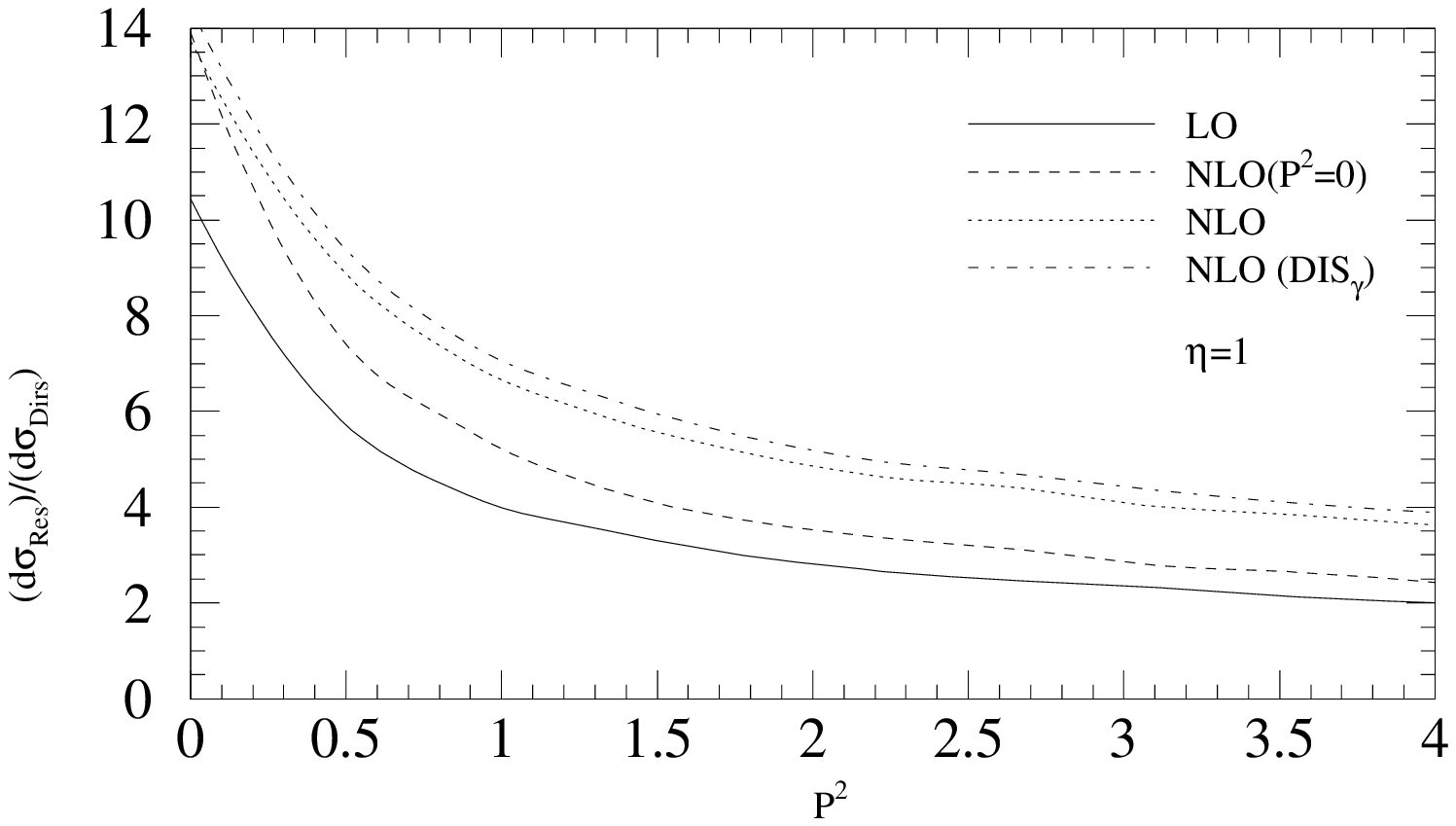,width=17cm}}
    \put(0,66){\parbox[t]{16cm}{\sloppy Figure 3a: The ratio of the
        resolved to the subtracted direct contribution in LO and NLO
        for the GRS parametrization, both in the
        $\overline{\mbox{MS}}$- and the DIS$_\gamma$-scheme for
        $\eta=2$ and $E_T=7$ GeV. The dashed curve is for the
        NLO-matrix elements with $P^2=0$ for comparison.}} 
    \put(0,-54){\parbox[t]{16cm}{\sloppy Figure 3b: Same as figure 3a
        with $\eta =1$.}}
  \end{picture}
\end{figure}
\newpage

\begin{figure}[hhh]
  \unitlength1mm
  \begin{picture}(122,150)
    \put(-2,-20){\epsfig{file=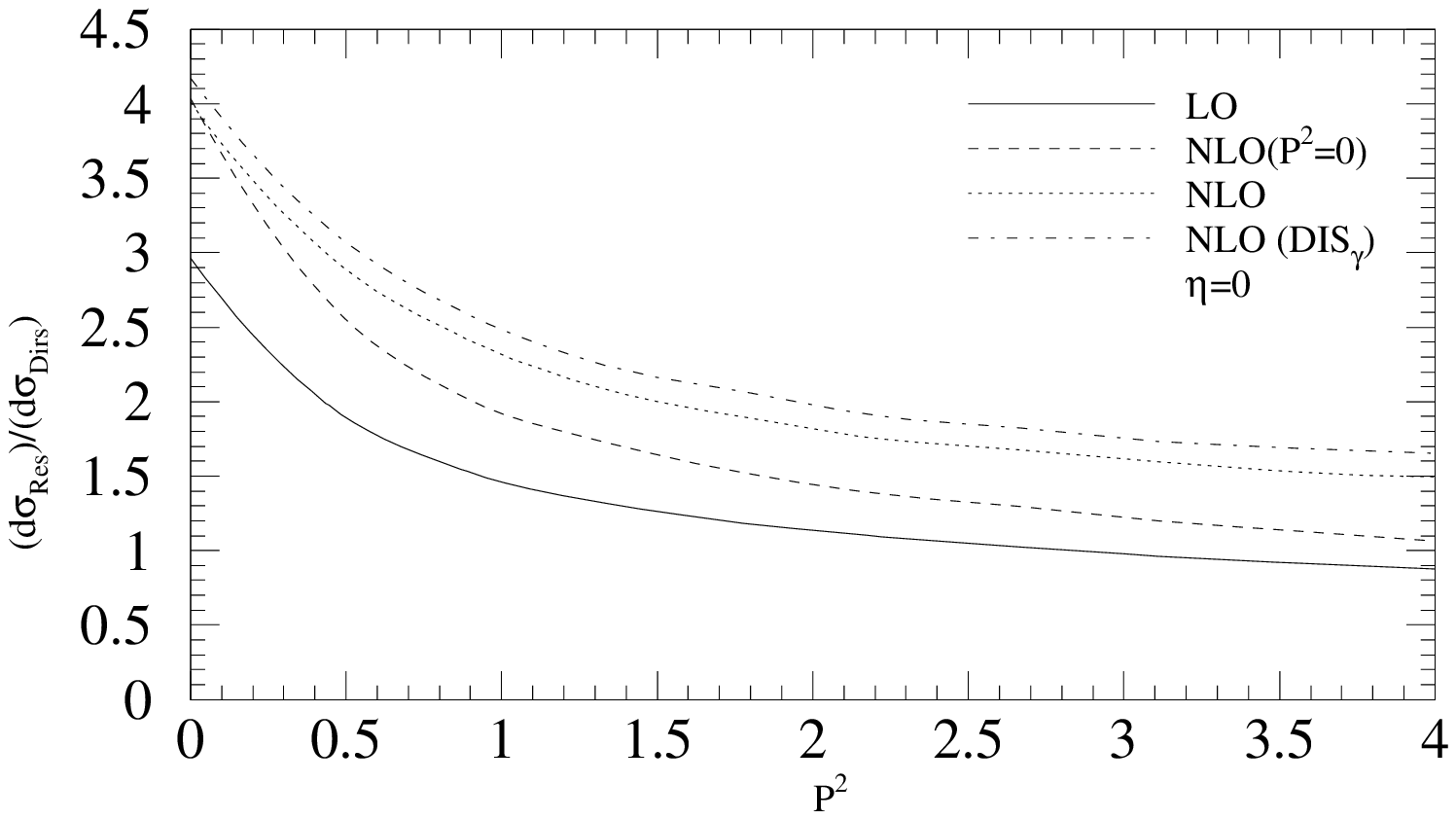,width=17cm}}
    \put(-2,-140){\epsfig{file=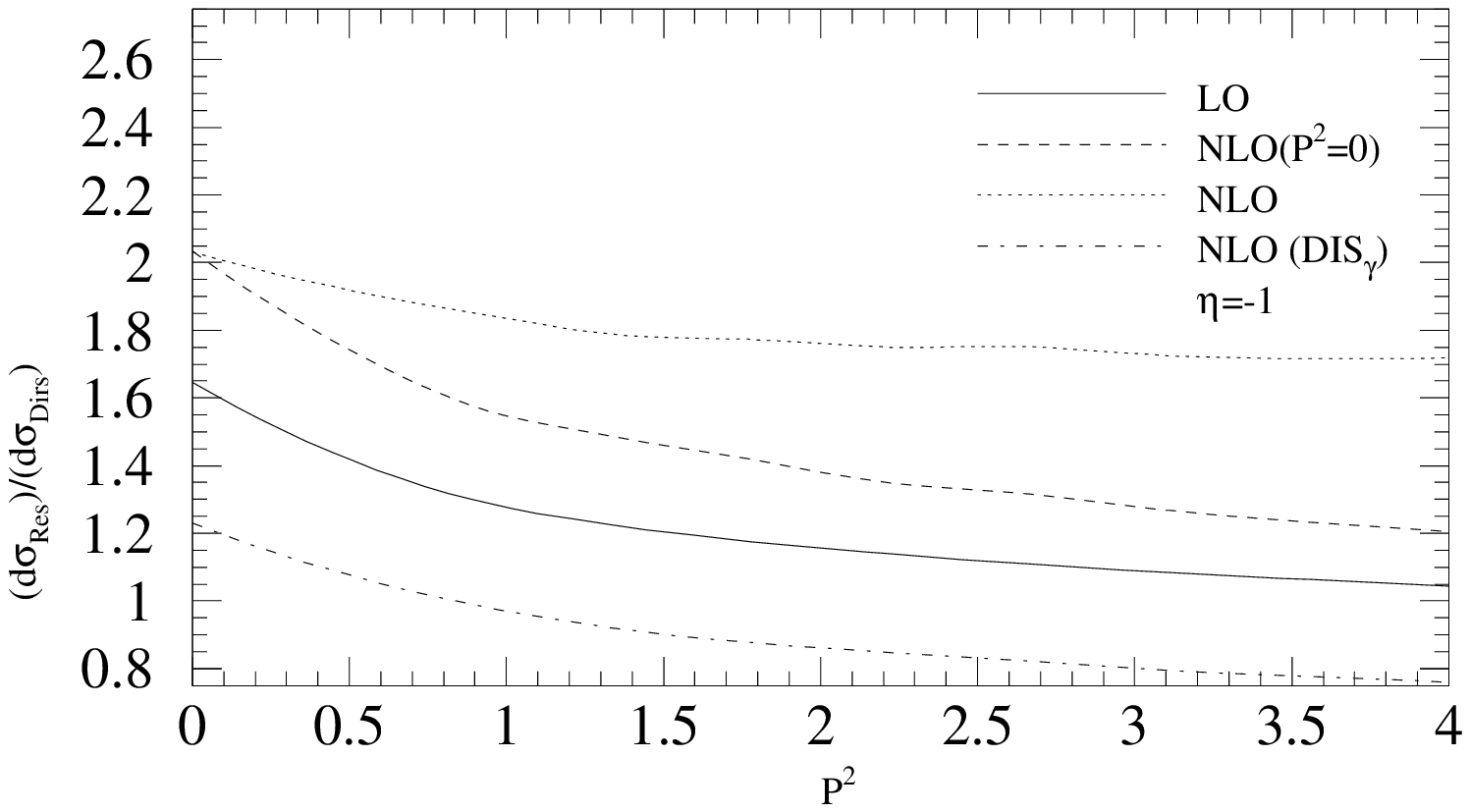,width=17cm}}
    \put(0,66){\parbox[t]{16cm}{\sloppy Figure 3c: Same as figure 3a
        with $\eta =0$.}}
    \put(0,-54){\parbox[t]{16cm}{\sloppy Figure 3d: Same as figure 3a
        with $\eta =-1$.}}
  \end{picture}
\end{figure}
\newpage

%*********************************************************************
% 4) 1-jet inclusive pt-spectrum, eta=[-1.125,1.875] (SaS)
%*********************************************************************

\begin{figure}[hhh]
  \unitlength1mm
  \begin{picture}(122,150)
    \put(-2,-20){\epsfig{file=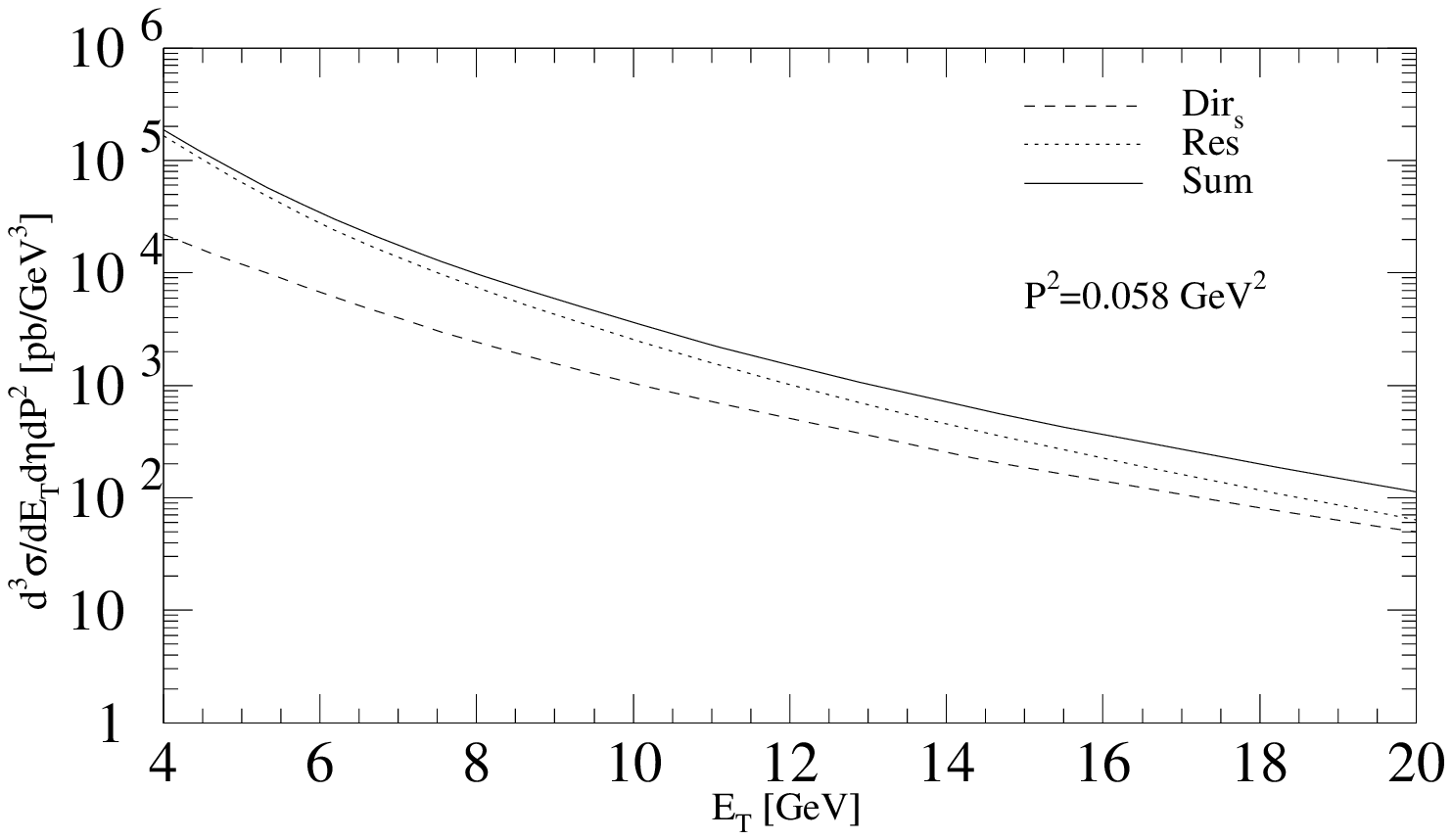,width=17cm}}
    \put(-2,-140){\epsfig{file=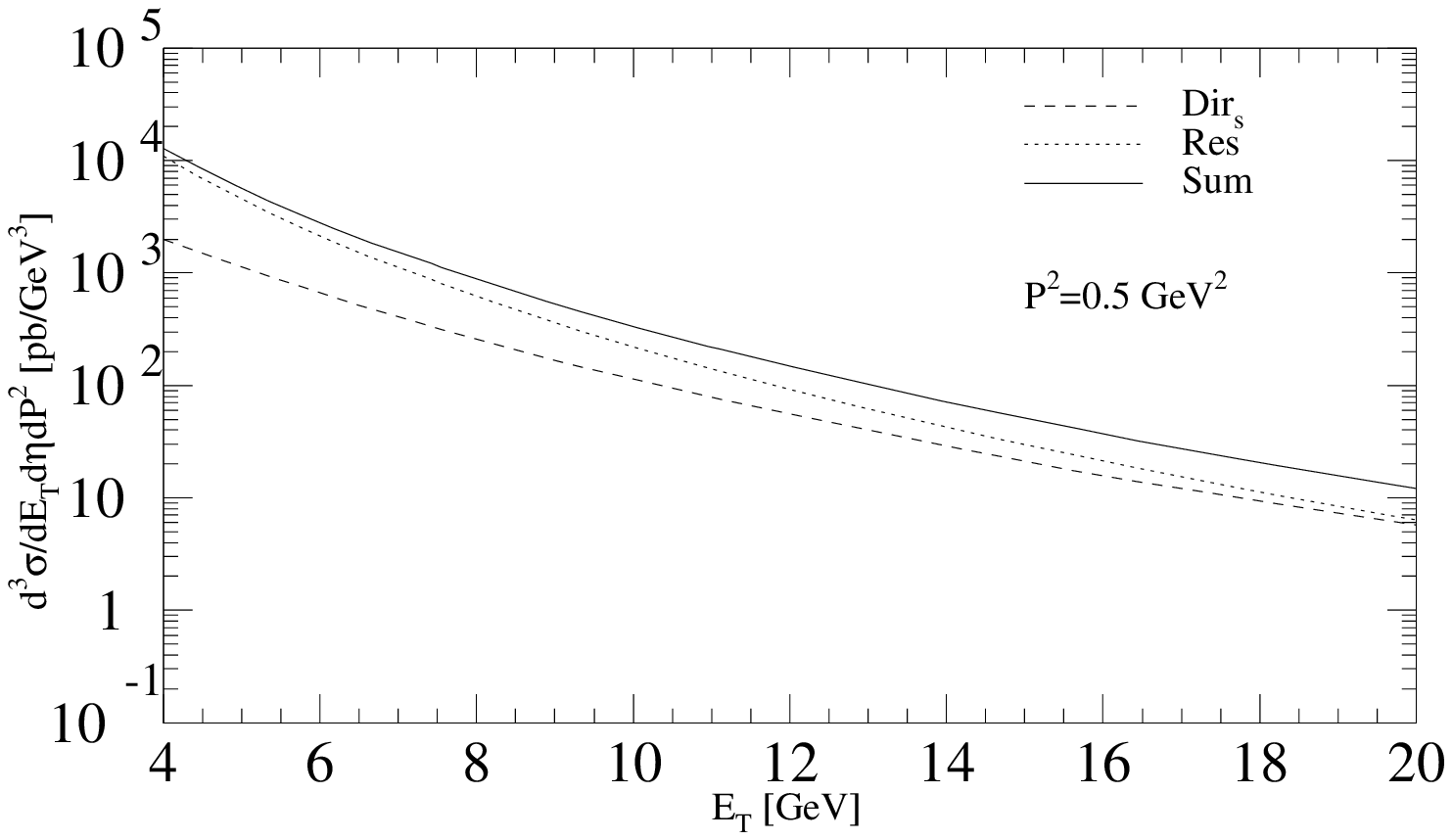,width=17cm}}
    \put(0,66){\parbox[t]{16cm}{\sloppy Figure 4a: Single-jet
        inclusive cross section integrated over $\eta \in
        [-1.125,1.875]$ for the virtuality $P^2=0.058$ GeV$^2$. The
        $\overline{\mbox{MS}}$-SaS1M
        parametrization with $N_f=4$ is chosen. The solid line gives
        the sum of the subtracted direct and the resolved term.}} 
    \put(0,-54){\parbox[t]{16cm}{\sloppy Figure 4b: Same as figure 4a
        with $P^2=0.5$ GeV$^2$.}}
  \end{picture}
\end{figure}
\newpage

\begin{figure}[hhh]
  \unitlength1mm
  \begin{picture}(122,150)
    \put(-2,-20){\epsfig{file=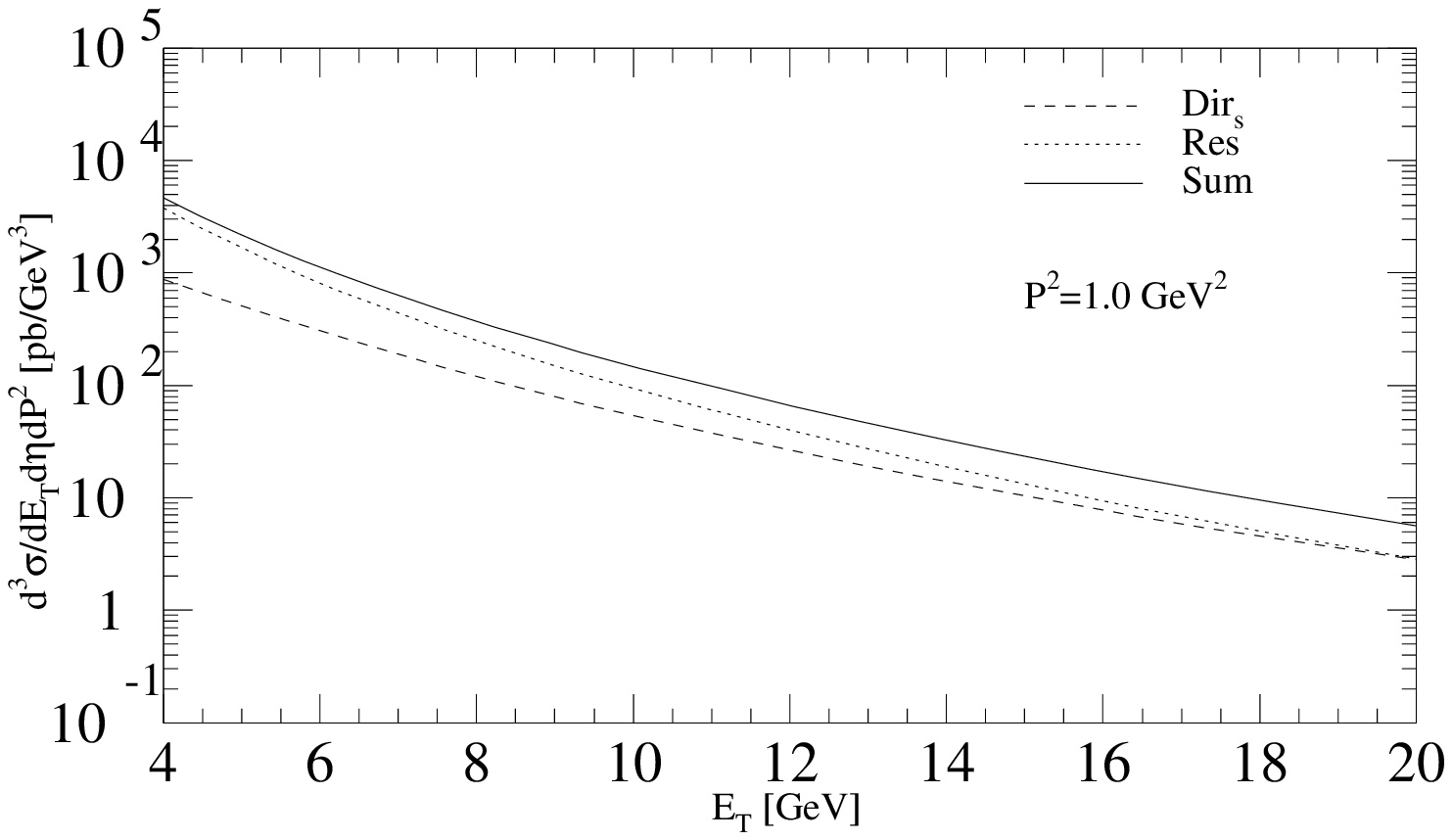,width=17cm}}
    \put(0,66){\parbox[t]{16cm}{\sloppy Figure 4c: Same as figure 4a
        with $P^2=1.0$ GeV$^2$.}}
  \end{picture}
\end{figure}
\newpage

%*********************************************************************
% 5) 1-jet inclusive eta-spectrum, pt=7
%*********************************************************************

\begin{figure}[hhh]
  \unitlength1mm
  \begin{picture}(122,150)
    \put(-2,-20){\epsfig{file=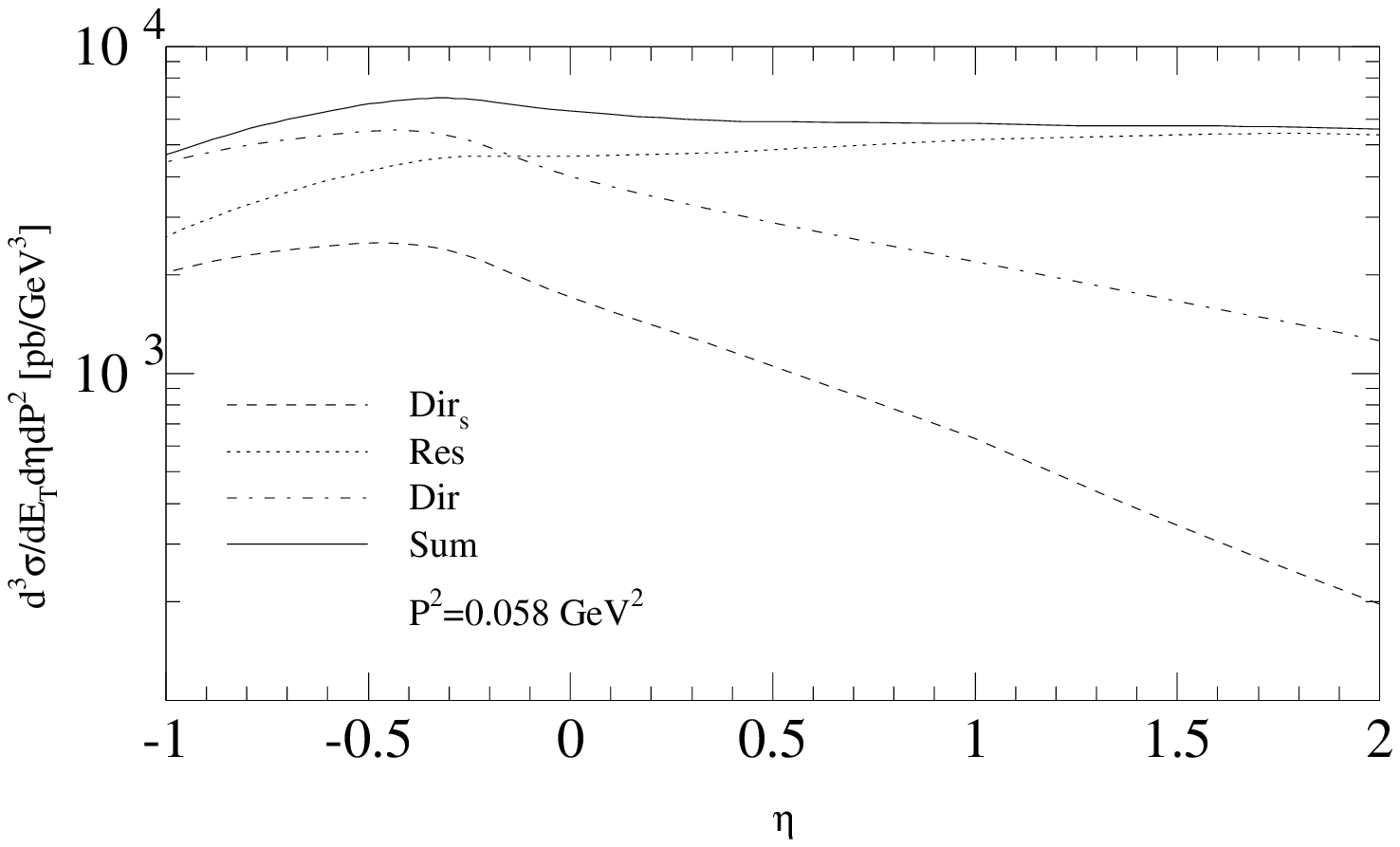,width=17cm}}
    \put(-2,-140){\epsfig{file=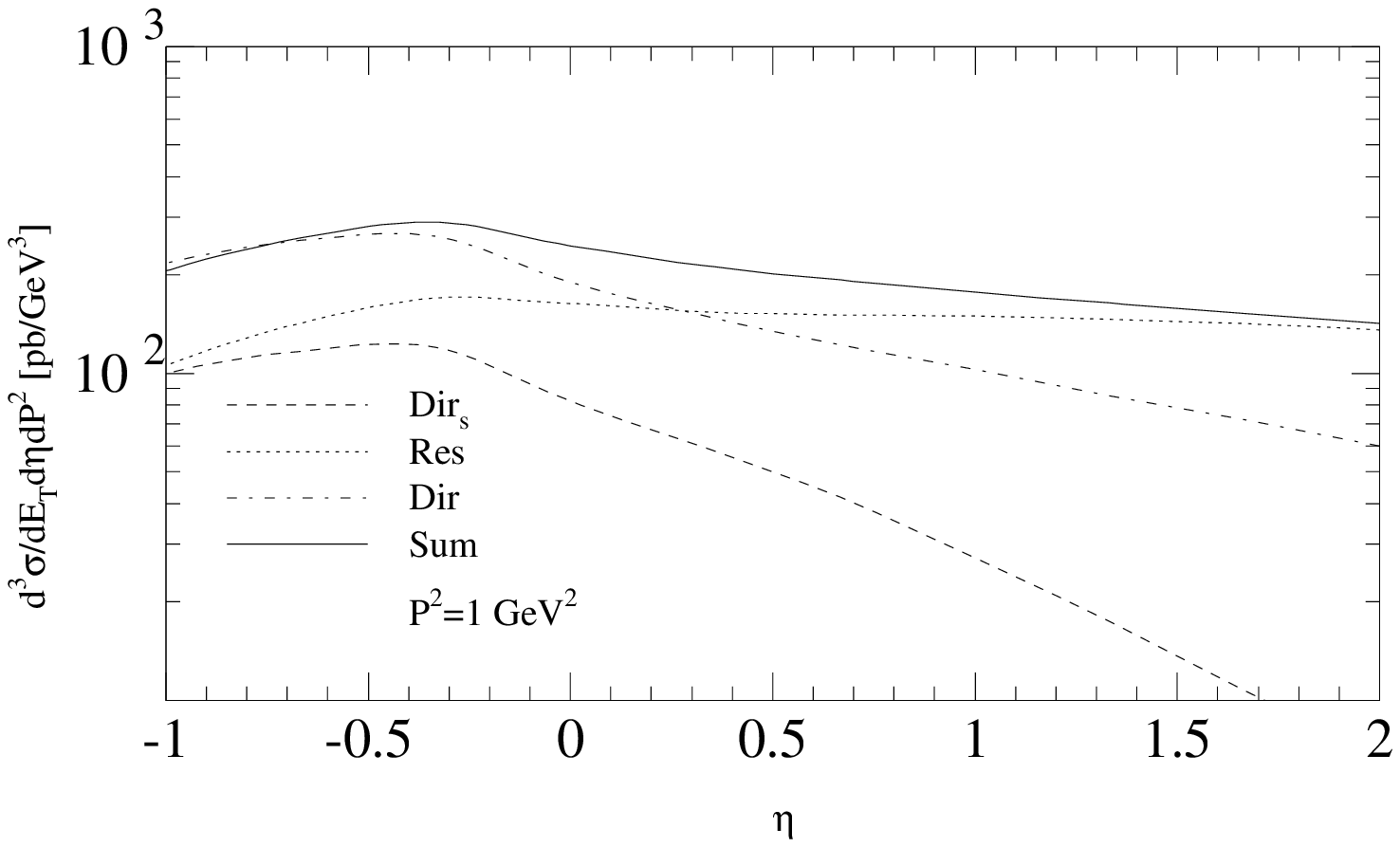,width=17cm}}
    \put(0,66){\parbox[t]{16cm}{\sloppy Figure 5a: Comparisons of
        single-jet inclusive cross sections for $E_T=7$ GeV and
        the virtuality $P^2=0.058$ GeV$^2$. The
        $\overline{\mbox{MS}}$-SaS1M parametrization with $N_f=4$ is 
        chosen. The solid line gives the sum of the subtracted direct
        and the resolved term. The dash dotted curve is the direct
        contribution without subtraction.}} 
    \put(0,-54){\parbox[t]{16cm}{\sloppy Figure 5b: Same as figure 5a
        with $P^2=1$ GeV$^2$.}}
  \end{picture}
\end{figure}
\newpage

\begin{figure}[hhh]
  \unitlength1mm
  \begin{picture}(122,150)
    \put(-2,-20){\epsfig{file=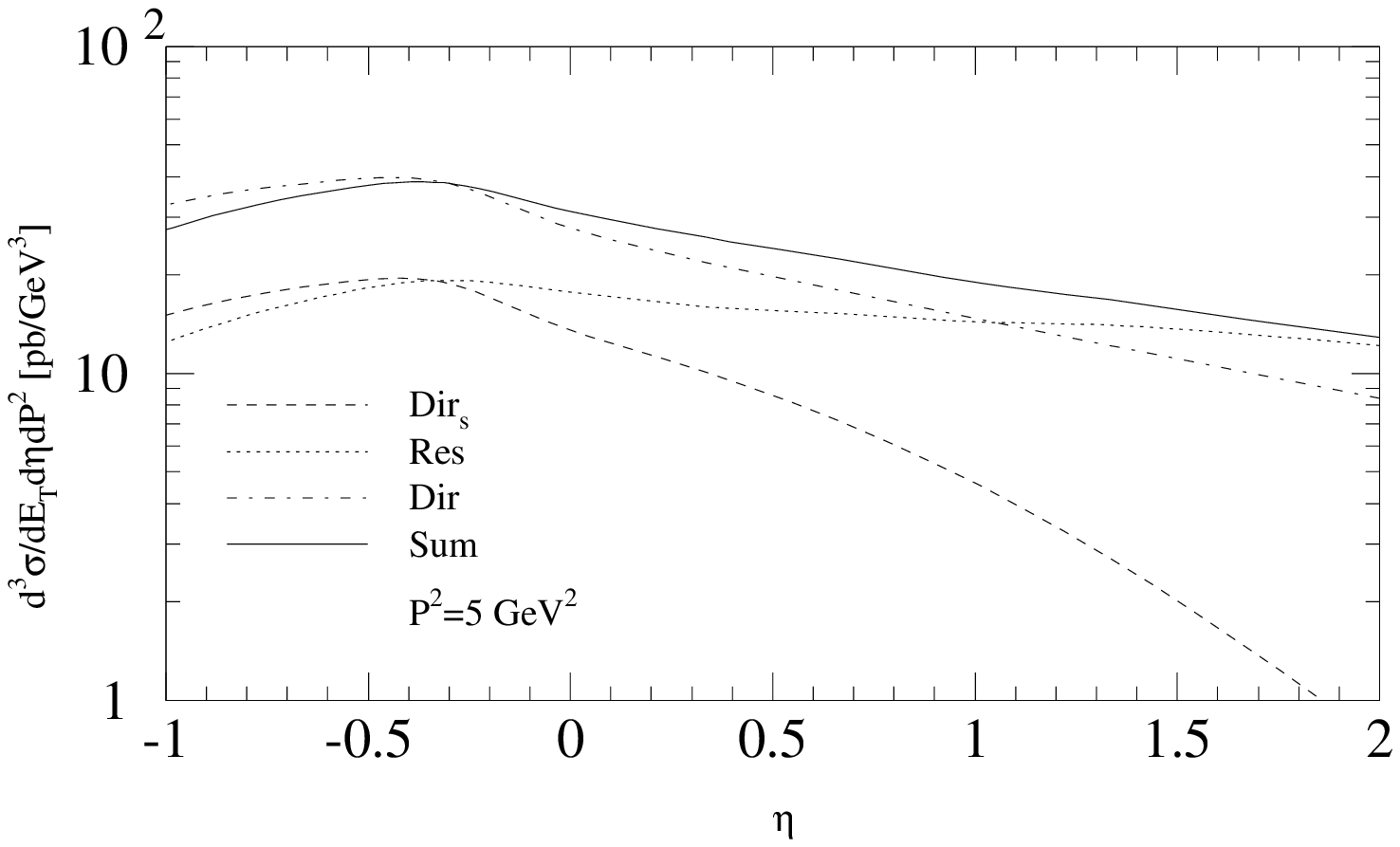,width=17cm}}
    \put(-2,-140){\epsfig{file=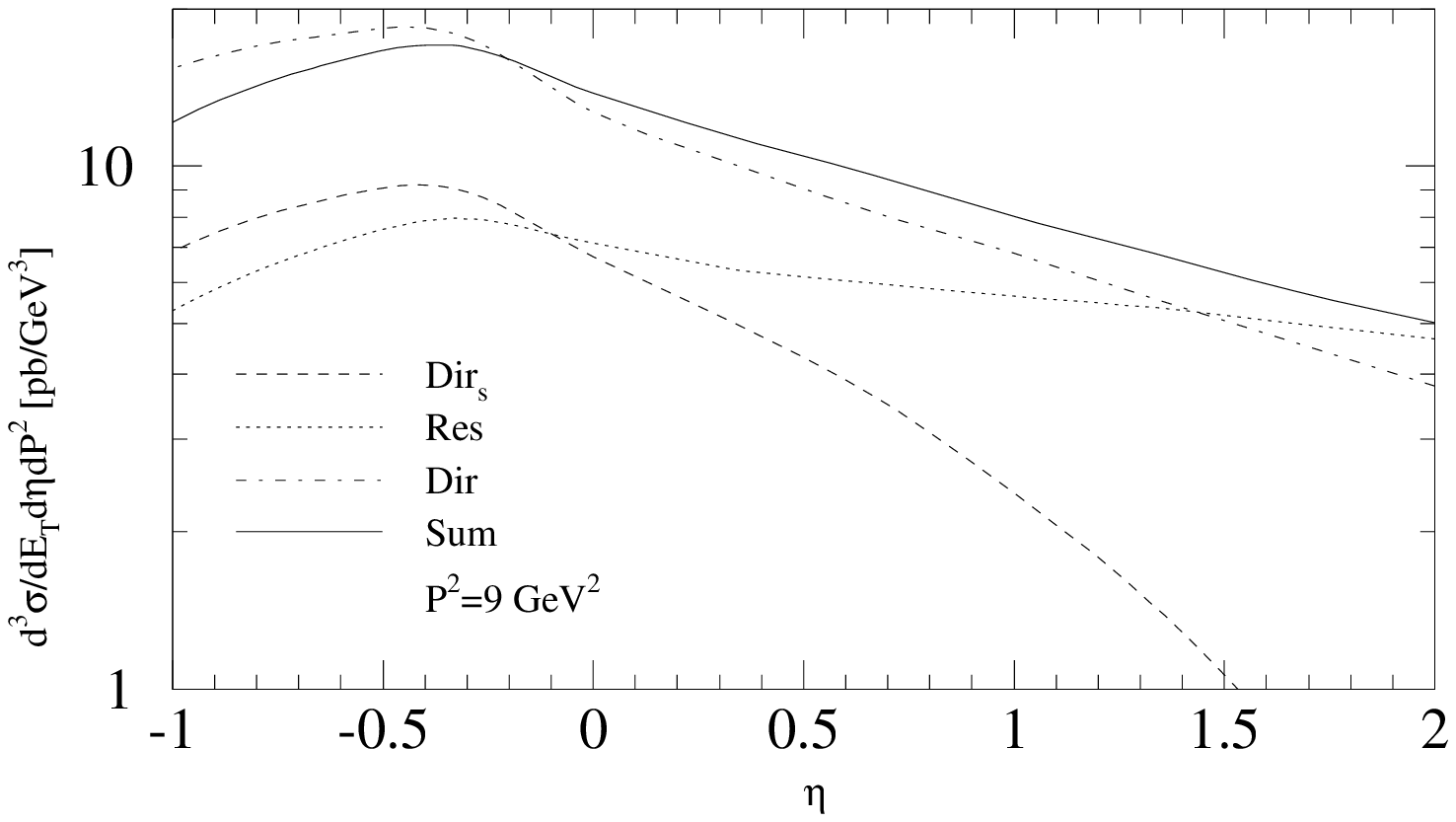,width=17cm}}
    \put(0,66){\parbox[t]{16cm}{\sloppy Figure 5c:  Same as figure 5a
        with $P^2=5$ GeV$^2$.}}
    \put(0,-54){\parbox[t]{16cm}{\sloppy Figure 5d: Same as figure 5a
        with $P^2=9$ GeV$^2$.}}
  \end{picture}
\end{figure}
\newpage

%*************************************************************************
% 6) und 7)  2-jet inclusive pt-spectra and P^2 spectrum (with ZEUS data)
%*************************************************************************

\begin{figure}[hhh]
  \unitlength1mm
  \begin{picture}(122,150)
    \put(-2,-20){\epsfig{file=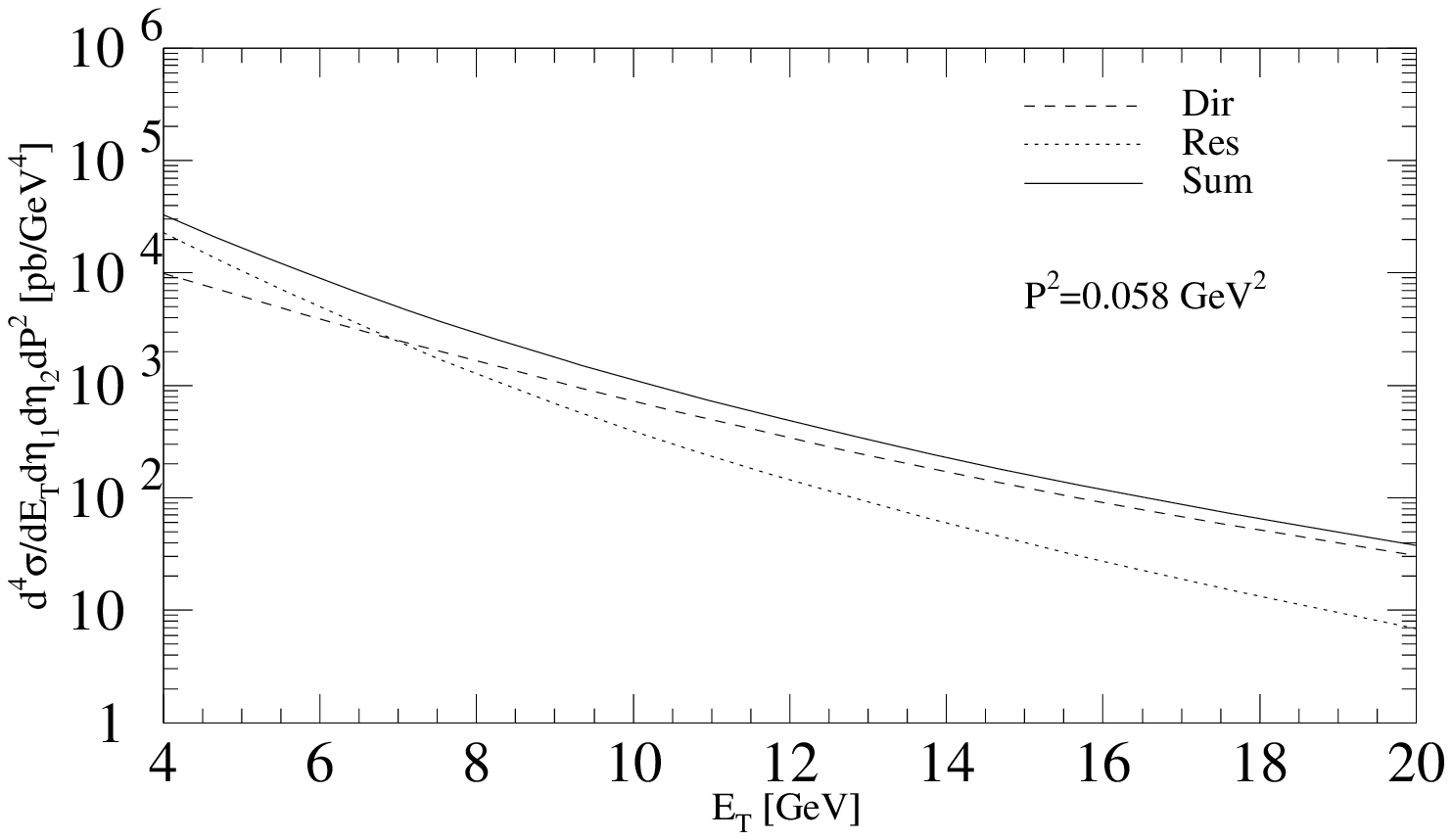,width=17cm}}
    \put(-2,-140){\epsfig{file=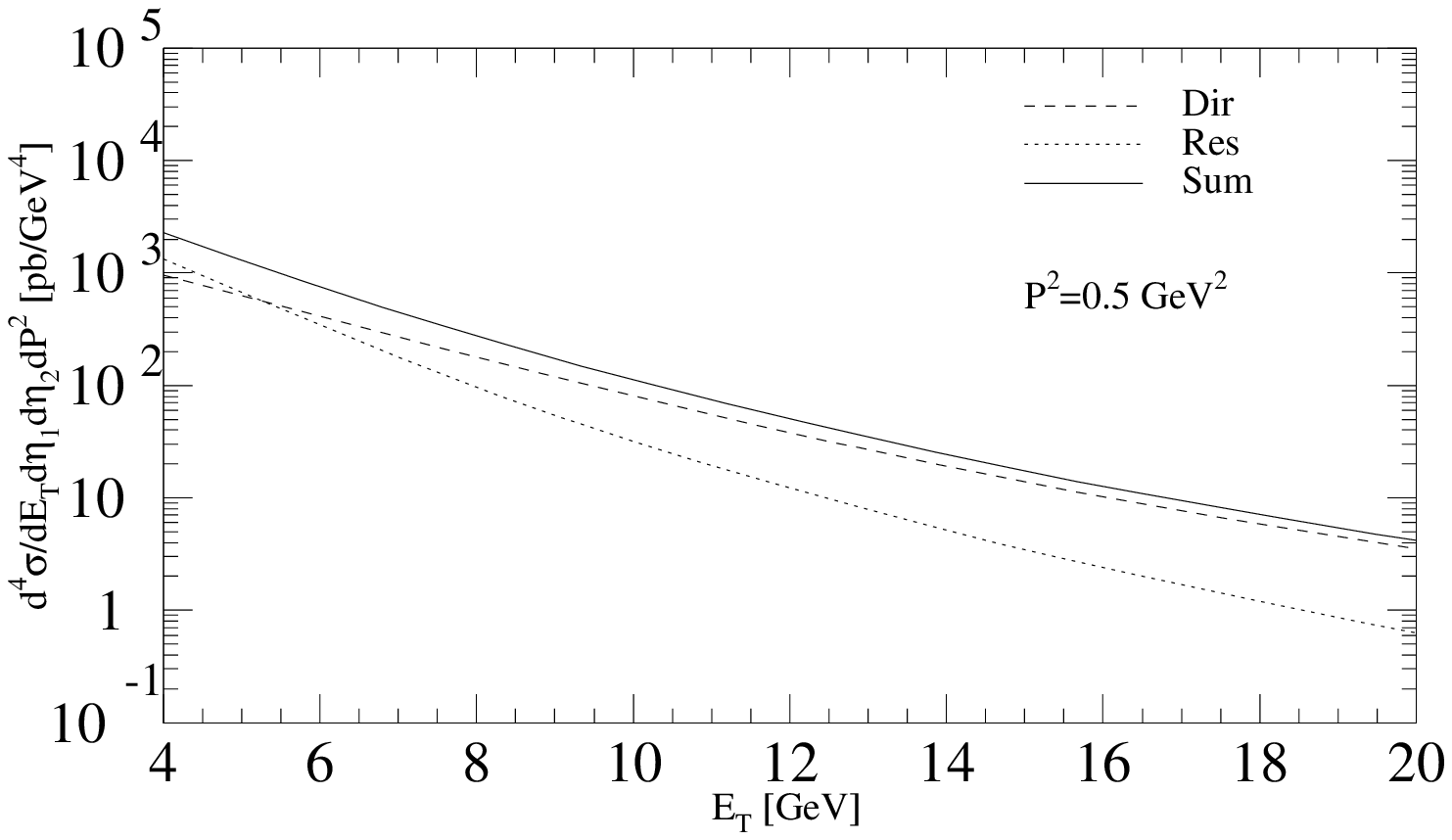,width=17cm}}
    \put(0,70){\parbox[t]{16cm}{\sloppy Figure 6a: Dijet inclusive
        cross section integrated over $\eta_1,\eta_2 \in
        [-1.125,1.875]$ for the virtuality $P^2=0.058$ GeV$^2$. The
        $\overline{\mbox{MS}}$-GRS parametrization with $N_f=3$ is
        chosen. The solid line is the sum of the direct and the
        resolved contribution. The dashed line is the direct-enriched
        contribution with $x_\gamma^{obs}>0.75$ and the dotted curve is
        the resolved enriched contribution with  $x_\gamma^{obs}<0.75$.}}
    \put(0,-54){\parbox[t]{16cm}{\sloppy Figure 6b: Same as figure 6a
        with $P^2=0.5$ GeV$^2$.}}
  \end{picture}
\end{figure}
\newpage

\begin{figure}[hhh]
  \unitlength1mm
  \begin{picture}(122,150)
    \put(-2,-20){\epsfig{file=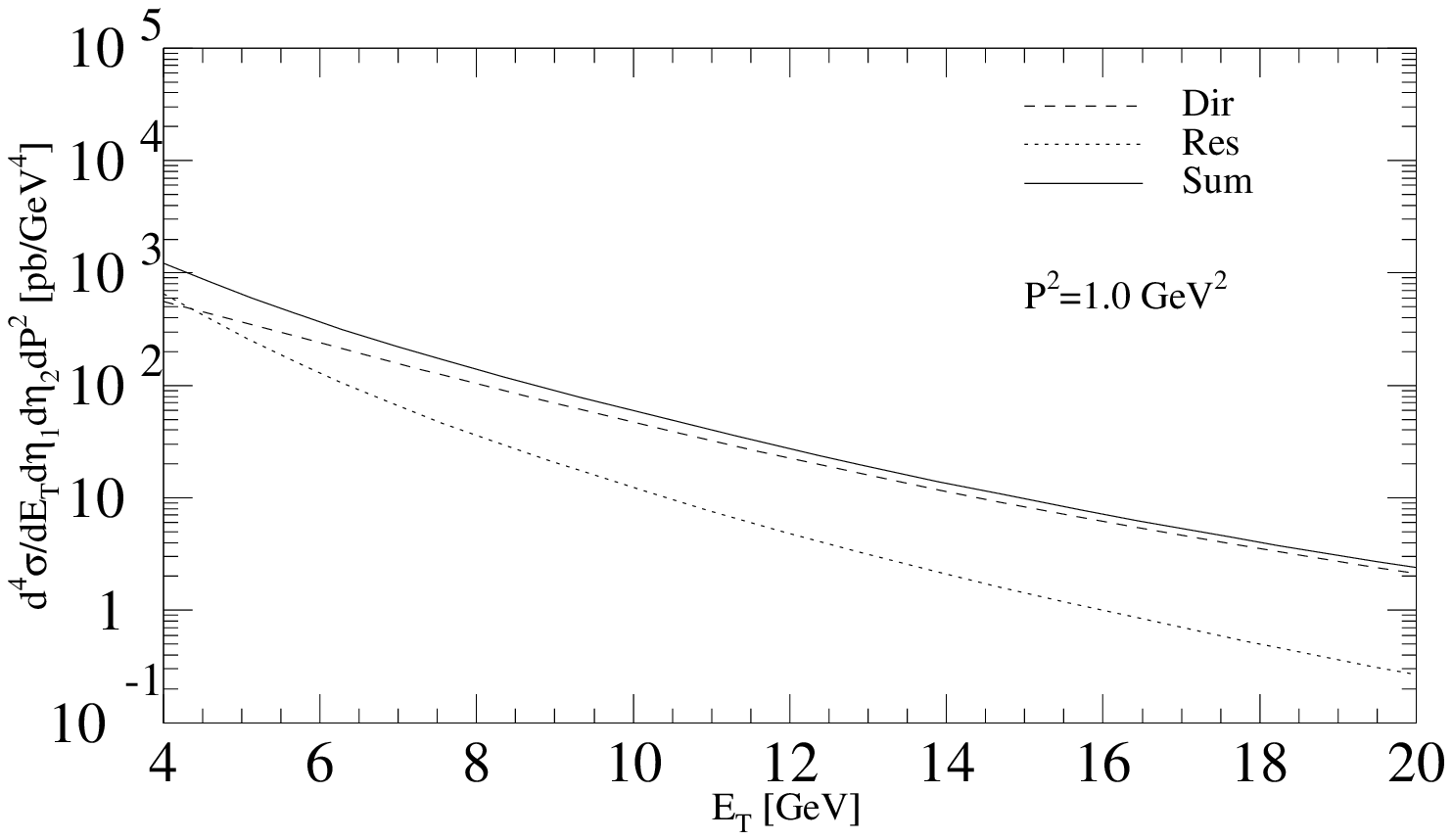,width=17cm}}
    \put(-2,-140){\epsfig{file=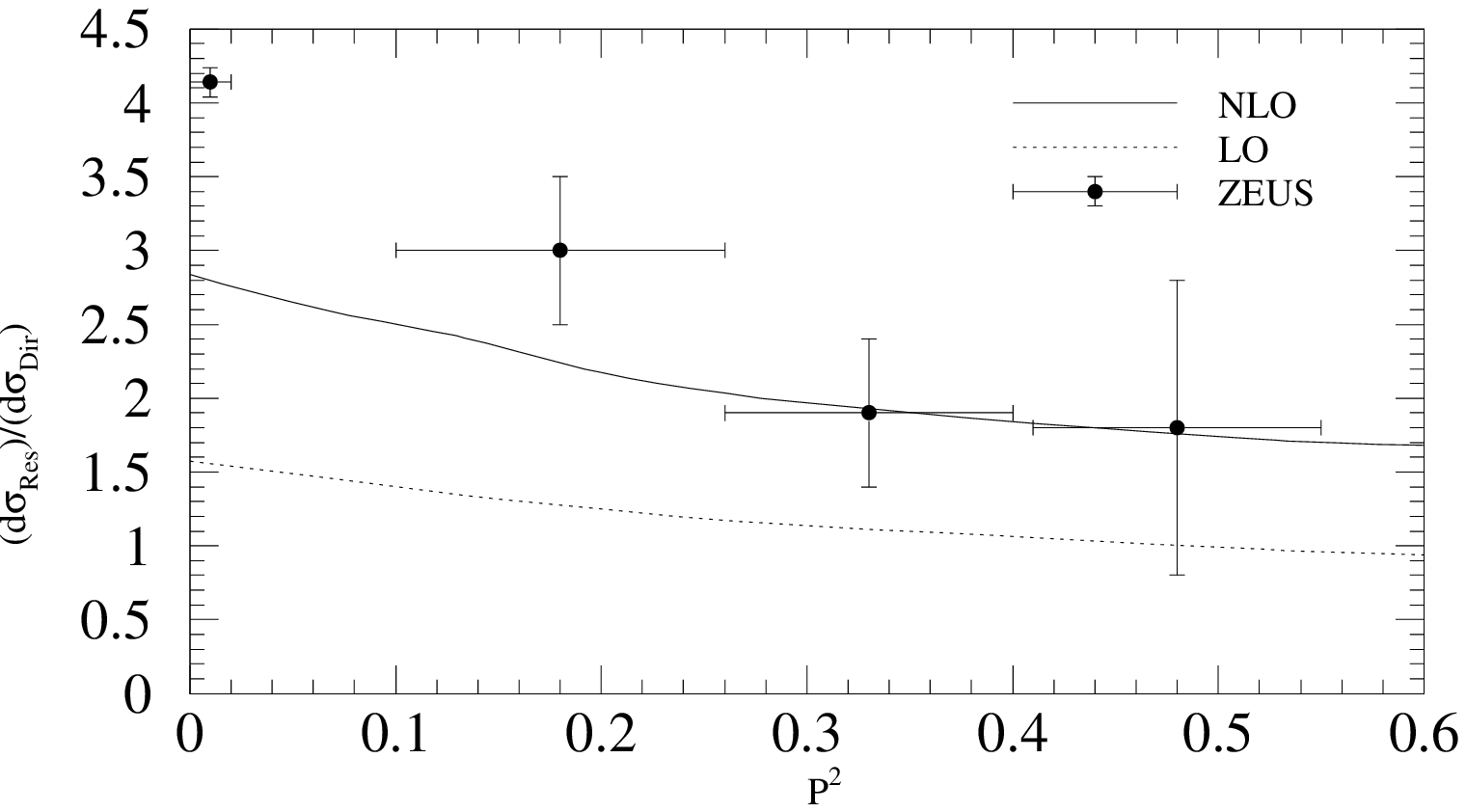,width=17cm}}
    \put(0,66){\parbox[t]{16cm}{\sloppy Figure 6c:  Same as figure 6a
        with $P^2=1.0$ GeV$^2$.}}
    \put(0,-54){\parbox[t]{16cm}{\sloppy Figure 7: The
        ratio of the resolved-enriched to the direct-enriched
        contributions as calculated in Fig.\ 6a, b, c, integrated over
        $E_{T_1},E_{T_2}>4$ GeV in LO (dotted) and NLO (full) for the
        SaS1M parametrization with $N_f=4$ compared
        with ZEUS data.}} 
  \end{picture}
\end{figure}

\end{document}